\documentclass{aa}

\usepackage{graphicx}
\usepackage{natbib}
\usepackage{scalerel}
\usepackage{amsmath}	
\usepackage{multicol}        
\usepackage{bm}		
\usepackage{pdflscape}	
\usepackage[usenames,dvipsnames,table]{xcolor}
\usepackage{tabularx}
\usepackage{multirow}
\usepackage{multicol}
\usepackage{verbatim}
\usepackage{tablefootnote}
\usepackage{threeparttable}

\newcommand{\cre}[1]{{\color{black}{#1}}}
\newcommand{\ros}[1]{{\color{black}{#1}}}

\usepackage{txfonts}
\usepackage[pdfencoding=auto,psdextra]{hyperref}
\hypersetup{
    colorlinks=true,
    linkcolor=blue,
    filecolor=magenta,      
    urlcolor=blue,
    citecolor=blue
}
\makeatletter
\renewcommand*\aa@pageof{, page \thepage{} of \pageref*{LastPage}}
\makeatother

\usepackage[utf8]{inputenc}

\usepackage[switch, modulo]{lineno}
\linenumbers

\def\Msun{\mbox{M$_\odot$}}

\def\mst{\mbox{$M_{\star}$}}

\def\lsim{\mathrel{\rlap{\lower3.5pt\hbox{\hskip0.5pt$\sim$}}
    \raise0.5pt\hbox{$<$}}}                
\def\gsim{~\rlap{$>$}{\lower 1.0ex\hbox{$\sim$}}}

\def\Fig{\mbox{Fig.~}}
\def\Figs{\mbox{Figs.~}}
\def\Tab{\mbox{Table~}}

\def\Sec{\mbox{Sect.~}}

\def\App{\mbox{Appendix~}}

\newcommand{\mulim}{$\mu_\mathrm{lim}$}

\def\Re{\mbox{$R_{\rm e}$}}

\def\sqd{\mbox{~sq.~deg.}}

\def\ic5332{\mbox{IC~5332}}
\def\hi{\mbox{H\,I}}
\def\ntwo{\mbox{[N\,II]}}
\newcommand{\ha}{$H{\alpha}$}
\DeclareUnicodeCharacter{2212}{-} 
\begin{document}
%
%

\title{VST-SMASH: VST Survey of Mass Assembly and Structural Hierarchy}
\subtitle{I. Survey presentation and deep photometry of \ic5332: tracing the mass assembly in the challenging faintest-end regime}
   
\newcommand{\orcid}[1]{} 

\author{
R.~Ragusa,$^{1}$\thanks{E-mail: rossella.ragusa@inaf.it}
C.~Tortora,$^{1}$
L.~Hunt,$^{2}$
M.~Spavone,$^{1}$
M.~Baes,$^{3}$
Abdurro'uf,$^{4}$
M.~Gatto,$^{1}$
F.~Annibali,$^{5}$
A.~Mercurio,$^{1,6,7}$
N.~Bellucco,$^{8}$
A.~Unni,$^{1}$
E.~Schinnerer$^{9}$ 
}
\institute{$^{1}$ INAF -- Osservatorio Astronomico di Capodimonte, Salita Moiariello 16, I-80131, Napoli, Italy\\
$^{2}$ INAF -- Osservatorio Astrofisico di Arcetri, Largo Enrico
Fermi 5, 50125, Firenze, Italy\\
$^{3}$ Sterrenkundig Observatorium, Universiteit Gent, Krijgslaan 281
S9, 9000 Gent, Belgium\\
$^{4}$ Department of Astronomy, Indiana University, 727 East Third Street, Bloomington, IN 47405, USA\\
$^{5}$ INAF - Osservatorio di Astrofisica e Scienza dello Spazio di Bologna, Via Gobetti 93/3, I-40129 Bologna, Italy\\
$^{6}$ Università di Salerno, Dipartimento di Fisica “E.R. Caianiello”, Via Giovanni Paolo II 132, 84084 Fisciano (SA), Italy.\\
$^{7}$ INFN – Gruppo Collegato di Salerno – Sezione di Napoli, Dipartimento di Fisica “E.R. Caianiello”, Università di Salerno, Via Giovanni Paolo II, 132, 84084 Fisciano (SA), Italy.\\
$^{8}$ Department of Physics and Astronomy, University of Padua, Italy\\
$^{9}$ Max Planck Institute for Astronomy, K\"unigstuhl  17, D-69117 Heidelberg, Germany\\
}

%
%
\abstract
   {Understanding the formation and evolution of late-type galaxies requires deep imaging for tracing the faintest stellar components in their outskirts (i.e., accreted material, diffuse halos, and tidal streams). Despite their crucial role in the build-up of stellar mass, these low surface brightness features remain largely unexplored due to observational limitations. The VST-SMASH (VST--Survey of Mass Assembly and Structural
Hierarchy) is designed to fill this gap, providing deep, wide-field optical imaging for a volume-limited sample of nearby spiral galaxies ($D<11$ Mpc), overlapping with the {\it Euclid} Wide Survey in the Southern Hemisphere.}
   {This paper aims to introduce the VST-SMASH survey and showcase its scientific potential through the analysis of \ic5332, a late-type spiral galaxy observed in the $g$, $r$, and $i$ bands. The main goal is to demonstrate the depth, quality, and diagnostic power of the dataset in tracing low-surface brightness features and structural components in galactic outskirts.

}
   {We carried out detailed surface photometry of \ic5332, using elliptical isophotal fitting and growth curve analysis to extract radial surface brightness and color profiles down to $\mu_r \sim 30$ mag arcsec$^{-2}$. We performed multi-component Sérsic decompositions in each band and constructed stellar mass surface density profiles. Furthermore, we identified and characterized faint stellar streams visible in the very deep and smoothed images, estimating their colors and comparing them with adjacent galactic regions.

}
   {While the internal ($<$ 1 effective radius) negative colour gradients can be explained by dissipative (monolithic) collapses and supernovae outflows, the color profiles at larger radii reveal 
   a significant gradient toward redder $g-r$ and $g-i$, spatially consistent with the presence of accreted populations in the outskirts. Nevertheless, we also find bluer $r-i$, which , togheter with the behaviour of $g-r$ and $g-i$, could be explained by strong \ha\ emission. Single-component Sérsic fits fail to reproduce the observed light distribution, especially in the outer regions, requiring a two-component model.
   %
   %
   In the outskirts, multiple faint stellar streams are detected, with integrated $g - r$ colors consistent with disrupted dwarf and/or low surface brightness satellites. However, spectroscopic follow-ups would be needed to detect possible strong emission lines, which could affect the $r$-band photometry. These findings support a scenario of ongoing stellar mass assembly through accretion and highlight the capability of VST-SMASH to uncover faint structures in nearby galaxies.

}
   {}

%
%
    
   \titlerunning{VST-SMASH - I: deep photometry of \ic5332}
   \authorrunning{R. Ragusa}
   
   \maketitle
   \nolinenumbers
\section{Introduction}\label{sec:intro}

The study of the outer halos of galaxies and, in particular, of the low surface brightness (LSB) structures, is pivotal for understanding the hierarchical nature of galaxy formation and the complex evolutionary pathways that lead to the diversity observed in the modern universe. Theoretical frameworks grounded in the $\Lambda$CDM (Lambda Cold Dark Matter) cosmological model, the most accredited scenario, posit that the universe's structure develops hierarchically: small-scale perturbations in the early universe collapse under gravity to form the first galaxies, which then merge and accrete material over time to create larger, more complex systems \citep[e.g.,][]{Blumenthal1984Natur.311..517B,WhiteFrenk1991ApJ...379...52W,Springel2008MNRAS.391.1685S}.

Within this scenario, LSB structures such as tidal streams, stellar halos, diffuse plumes, and intra-group or intra-cluster light (IGL or ICL) preserve the fossil record of past gravitational interactions. These features typically arise from minor mergers \citep[mass ratios $\sim$1:10, or lower; e.g., ][]{Conselice2014}; the disruption of dwarf galaxy satellites, or the stripping of satellite outskirts.
Their origin can be investigated through radial colour profiles, tracing stellar population gradients from the galaxy core to its faint periphery. Mergers tend to flatten these gradients, whereas tidal stripping or dwarf disruption produces stronger colour variations \citep[e.g.,][]{montes2022NatAs...6..308M}.
Deep, wide-field, and multi-band photometry is therefore essential to probe stellar populations and reveal the structure of faint outskirts. The resulting streams, arcs, and diffuse remnants also trace the underlying dark matter distribution \citep[e.g.,][]{johnston01,cooper2010MNRAS.406..744C,Nibauer+2025,Starkman+25}, and provide critical tests for numerical simulations of galaxy growth \citep{bell2008ApJ...680..295B,sandford2017MNRAS.470..522S}.

Detecting and characterizing LSB features is observationally challenging, as they typically lie below $\mu_V \sim 27$ mag arcsec$^{-2}$. Such studies require wide-field, deep, and high-resolution imaging to distinguish faint emission from sky background, scattered light, or Galactic cirrus, as well as the light coming from foreground/background sources
in the field \citep[e.g.][]{trujillo16,roman20,ragusa2021A&A...651A..39R}.
Traditional surveys have been constrained by sky brightness, detector sensitivity, and limited field of view (FoV), preventing the recovery of extended diffuse structures \citep{martinez10}. Even space-based facilities such as the Hubble Space Telescope (HST), despite their resolution, are hindered by a narrow FoV that restricts mapping of galactic outskirts \citep{radburn11}.

Until recently, detecting LSB features was extremely challenging. The advent of Gaia and the Dark Energy Survey (DES) has marked a ‘golden age’ for stellar stream studies \citep[e.g.][]{Belokurov2017MNRAS.466.4711B,shipp2018ApJ...862..114S}. While these surveys have provided key results within the Local Group (LG), deeper imaging beyond it is crucial to fully trace galaxy assembly.
So far, systematic searches for tidal structures outside the LG remain limited, emphasizing the need for wider, homogeneous studies. Ground-based surveys such as MATLAS (Mass Assembly of early-Type GaLAxies with their fine Structures; \citealt{bilek2020MNRAS.498.2138B}), ELVES (Exploration of Local VolumE Satellites; \citealt{carlsten22}), SSH (Smallest Scale of Hierarchy Survey; \citealt{annibali2020MNRAS.491.5101A}), and LIGHTS (LBT Imaging of Galactic Halos and Tidal Structures; \citealt{Trujillo2021A&A...654A..40T})
have significantly advanced the detection of faint stellar streams and satellites, though most focus on the northern sky. The southern hemisphere remains underexplored—an important gap given its overlap with major upcoming surveys such as {\it Euclid}.
The VLT Survey Telescope (VST) at Paranal Observatory (Chile) is uniquely suited to bridge this gap. Its OmegaCAM, with a 1-square-degree FoV, is ideal for deep imaging across extensive areas \citep{Arnaboldi1998Msngr..93...30A,Kuijken2002Msngr.110...15K}. The VST has shown its capability through successful projects like VEGAS \citep[VST Early-type Galaxy Survey;][]{Capaccioli2015A&A...581A..10C} and the Fornax Deep Survey \citep[FDS;][]{Iodice2016ApJ...820...42I}. These surveys demonstrated that VST can achieve surface brightness (SB) limits down to $\sim$ 29–30 mag/arcsec$^2$ in the $g$ and $r$ bands, which are deep enough to identify diffuse LSB features of both early-type and late-type galaxies \citep[e.g.,][]{Iodice2016ApJ...820...42I,IODICE2017ApJ...851...75I, spavone2017Galax...5...31S,ragusa2021A&A...651A..39R,ragusa2022FrASS...952810R,ragusa2023A&A...670L..20R}.

The VST Survey of Mass Assembly and Structural Hierarchy (VST-SMASH, P.I. C. Tortora) aims to leverage these strengths to conduct a systematic survey of 27 nearby galaxies (D $\leq$ 11 Mpc) within the {\it Euclid} footprint \citep{Tortora+24_VST-SMASH}. This survey, conducted in the $g$, $r$, and $i$ bands, offers a unique opportunity to map LSB features and probe the mass assembly history of galaxies in the Local Volume. By extending the depth of imaging and capturing the halo structures, VST-SMASH can reveal subtle, yet critical, components of galaxy assembly, including stellar streams and outer shells, which are often missed by shallower surveys. Beyond morphology, LSB observations constrain the age and metallicity distributions of stellar populations in the outskirts \citep[e.g.][]{Sanderson2018ApJ...869...12S}, offering clues to galaxies’ accretion histories and merger processes. They also probe the underlying dark matter (DM) potential, sensitive to halo substructures predicted by $\Lambda$CDM models \citep[e.g.,][]{bell2008ApJ...680..295B,sandford2017MNRAS.470..522S,tulin2018PhR...730....1T,Nibauer+2025,Starkman+25}, thus linking baryonic assembly to DM structure formation.
Observatories such as the Rubin Observatory’s Legacy Survey of Space and Time (LSST) and {\it Euclid} will greatly advance LSB studies thanks to their wide FoV and deep imaging capabilities, though their full impact will emerge only over the next decade. LSST will reach the required SB limits after several years of cumulative observations \citep[$\mu_r \sim 30.3$ mag arcsec$^{-2}$ in $\sim$10 years;][]{Brough2024MNRAS.528..771B}, while {\it Euclid}, operating mainly in the NIR with a single broad optical filter (VIS), will lack the full colour information needed for detailed LSB analyses.
In this context, VST–SMASH can deliver immediate scientific returns and acts as an optical complement to {\it Euclid}. Providing deep, multi-band, wide-field optical imaging of southern-sky galaxies, it fills a crucial observational gap and complements other existing surveys such as MATLAS, ELVES, SSH, LIGHTS, VEGAS, and FDS. Its results will yield essential datasets to test hierarchical formation models, constrain DM properties, and refine our understanding of galaxy evolution.

In this first paper, we: i) introduce the VST-SMASH survey, presenting the galaxy sample, observations, and data reduction, and verify that the planned survey depth is reached (\Sec\ref{sec:survey}); ii) illustrate the first VST-SMASH galaxy observed and reduced \ic5332, a low-mass, almost face-on spiral galaxy at a distance of 7.8 Mpc \citep{Meyer+04_HIPASS,Karachentsev+13}, together with the literature (\Sec\ref{sec:targets}); iii) present the deep and wide field photometric analysis for \ic5332 (\Sec\ref{sec:SB_and_colour_profiles}); iv) discuss the results  (\Sec\ref{sec:stream}), and draw conclusions, and future prospects  (\Sec\ref{sec:discussion_conclusione}). 
The magnitudes throughout the paper are provided in the AB system, and are corrected for Galactic extinction using the dust maps of \citet{Schlafly_2011}. The $V$-band foreground extinction\footnote{We retrieved foreground extinction in the three bands from the NASA/IPAC Extragalactic Database (NED, https://ned.ipac.caltech.edu).}. is $A_{\rm V} = 0.046$ mag, which corresponds to $A_{\rm g} = 0.063$ mag, $A_{\rm r} = 0.046$ mag, and $A_{\rm i} = 0.035$ mag in the VST $g$, $r$, and $i$ filters, respectively.

\begin{figure}
\centering
\includegraphics[width=0.85\linewidth]{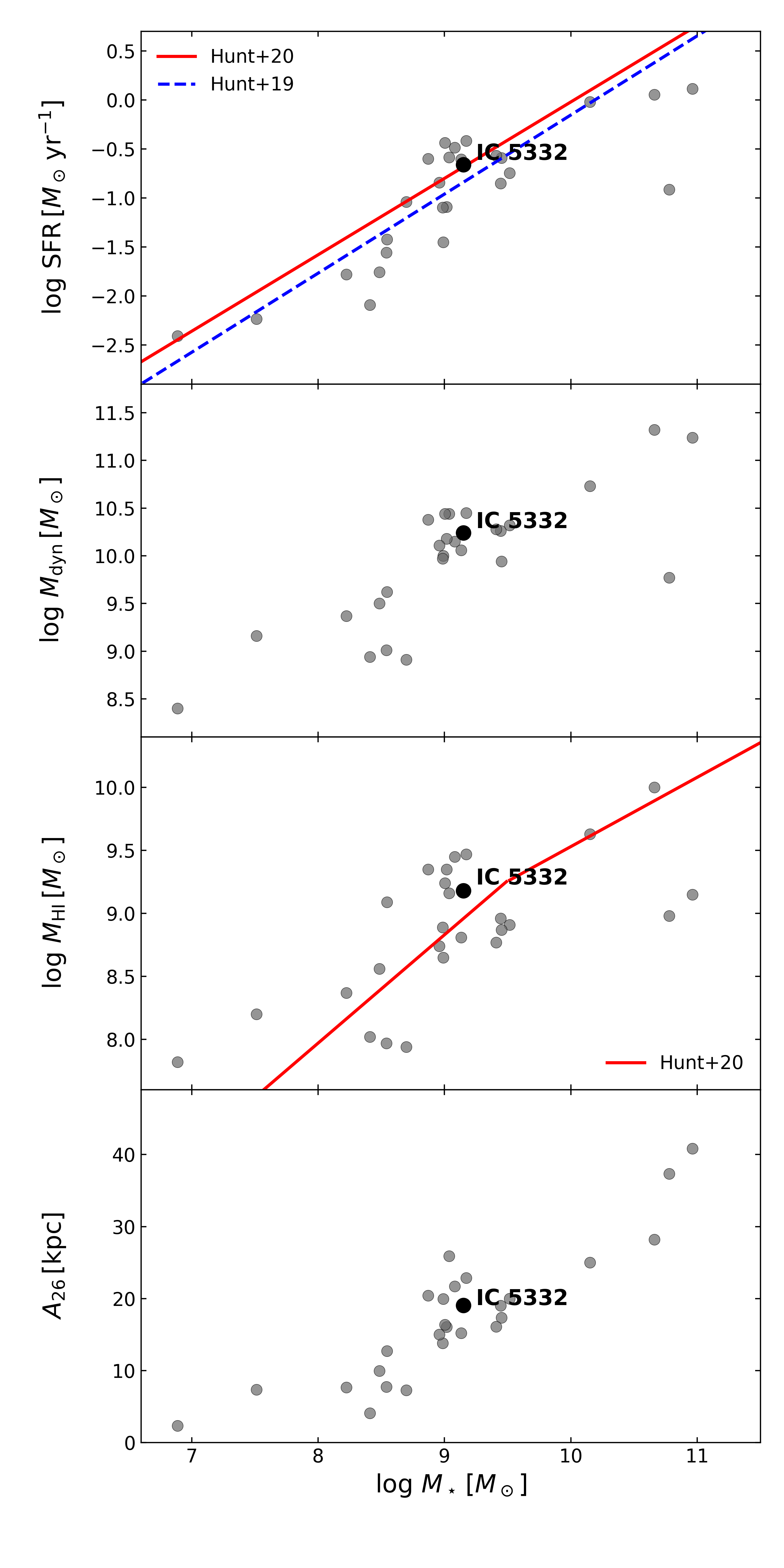}
\caption{\cre{From top to bottom, the panels show the SFR, dynamical mass, \hi\ gas mass, and Holmberg radius (in kpc) as functions of stellar mass for the VST–SMASH sample. The target analyzed in this paper, \ic5332, is highlighted with black symbols. SFR- and $M_{\rm HI}$-\mst\ best-fit relations from \citet{hunt19} and \citet{Hunt2020} are also shown.} }\label{fig:VST-SMASH_sample_properties}
\end{figure}

\section{The VST-SMASH survey}\label{sec:survey}

The primary goal of the VST-SMASH Survey is to achieve an extensive characterization of stellar populations, tidal structures, dwarf galaxies, globular clusters, and all the LSB features around a galaxy sample within a volume-limited distance of $\sim$ 11 Mpc. Leveraging the capabilities of the VST and OmegaCAM, the survey targets SB limits of $\sim$ 30, 30, and 28 mag/arcsec$^2$ in $g$, $r$, and $i$ bands in the AB system, respectively. This section outlines the sample selection defining the complete VST-SMASH survey (\Sec\ref{subsec:sample}), followed by a discussion on observation strategies and data reduction (\Sec\ref{subsec:obs_reduction}) and on the method adopted for determining the surface photometry (\Sec\ref{sec:SB_photometry_method}).

\subsection{Galaxy sample}\label{subsec:sample}

To comprehensively characterize mass assembly processes within a distance of 11 Mpc, we selected a volume-limited galaxy sample that includes diverse objects in terms of morphological type, mass, and size, fully exploiting the capabilities of the VST and OmegaCAM. By observing both the target fields and an adjacent field for effective background subtraction (see \Sec\ref{subsec:obs_reduction} for a detailed description), we can analyze the galaxies' halos up to roughly 250 kpc, depending on the targets’ distances.
The sample was chosen from the Updated Nearby Galaxy Catalogue \citep[UNGC,][]{Karachentsev+13}, which includes 869 galaxies within 11 Mpc, applying the following criteria:
\begin{itemize}
\item Declination: only galaxies with declinations of $\leq$ 5 $\deg$ were included to ensure optimal visibility conditions.
\item Holmberg diameter: we selected galaxies with Holmberg diameter (i.e., the radius at which $\mu_B$ = 26.5 mag arcsec$^{-2}$)  $A_{26} \geq 5$ arcmin to maximize the use of VST's wide FoV.
\item {\it Euclid} footprint: the targets were constrained to lie within the {\it Euclid} 
observational footprint.
\item Exclusions: i) large galaxies, such as the LMC, SMC, and Sag dSph, as well as Milky Way companions, were excluded due to extensive previous study; ii) NGC 3115 was excluded as it was already observed with VST; iii) galaxies heavily affected by overlapping or very bright stars in the field were also excluded.
\end{itemize}

This selection yielded a final sample of 27 galaxies, covering a broad range of properties. The galaxies span total masses between  $10^9 - 2\times 10^{11} \, \Msun$, stellar masses in the range $10^{7} - 10^{11} \, \Msun$,  \hi\ gas masses from $10^{8}$ to $10^{10} \, \Msun$, and a variety of morphological types, orientations, and environments. Their apparent sizes range significantly, with Holmberg diameters $A_{26}$ between 5 and 40 arcmin (corresponding to $2 - 40 \, \rm kpc$). A table summarizing the targets and their main properties, along with additional details on the survey objectives, is provided by \cite{Tortora+24_VST-SMASH}. In \Fig\ref{fig:VST-SMASH_sample_properties}, we present the main physical parameters of the VST–SMASH sample, showing how the star formation rate (SFR), the total dynamical mass, \hi\ gas mass, and Holmberg radius ($A_{\rm 26}$) correlate with stellar mass. The SFR is computed as the mean of the \ha- and FUV-based estimates from \citet{Karachentsev+13}, corrected to a Chabrier IMF by subtracting 0.25 dex \citep{Tortora+09AGN}. Dynamical masses within the Holmberg radius and \hi\ masses are taken from \citet{Karachentsev+13}, the former derived from rotation curves and the latter from integrated \hi\ fluxes. The Holmberg radius is converted to physical units using catalog distances. Stellar masses are derived from the catalog $BVK$ magnitudes using the mass-to-light ratio ($M/L$)–color relations of \citet{BdeJ01}, based on a scaled Salpeter IMF. The SFR- and $M_{\rm HI}$-\mst\ relations of the VST-SMASH sample are consistent with the literature \citep{hunt19,Hunt2020}.

In this work, we focus on \ic5332, intending to present preliminary results and the goals achieved by the survey. This galaxy will be described in \Sec\ref{subsec:IC5332}, with its characteristics outlined in \Tab\ref{tab:TARGET}, and its position in the various scaling relations shown in  \Fig\ref{fig:VST-SMASH_sample_properties}.

\begin{figure*}
\includegraphics[width=0.6\linewidth]{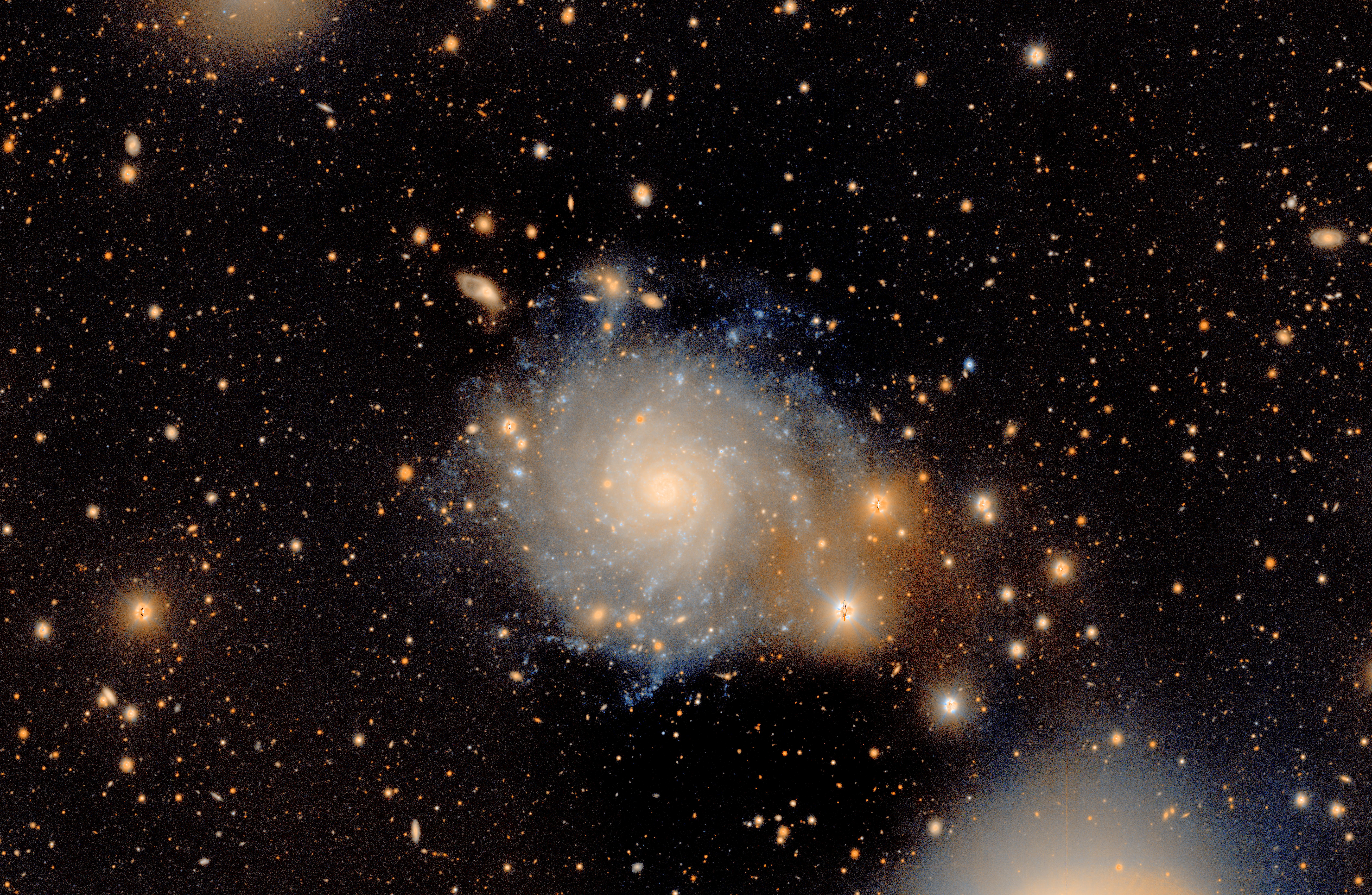}
\includegraphics[width=0.375\linewidth]{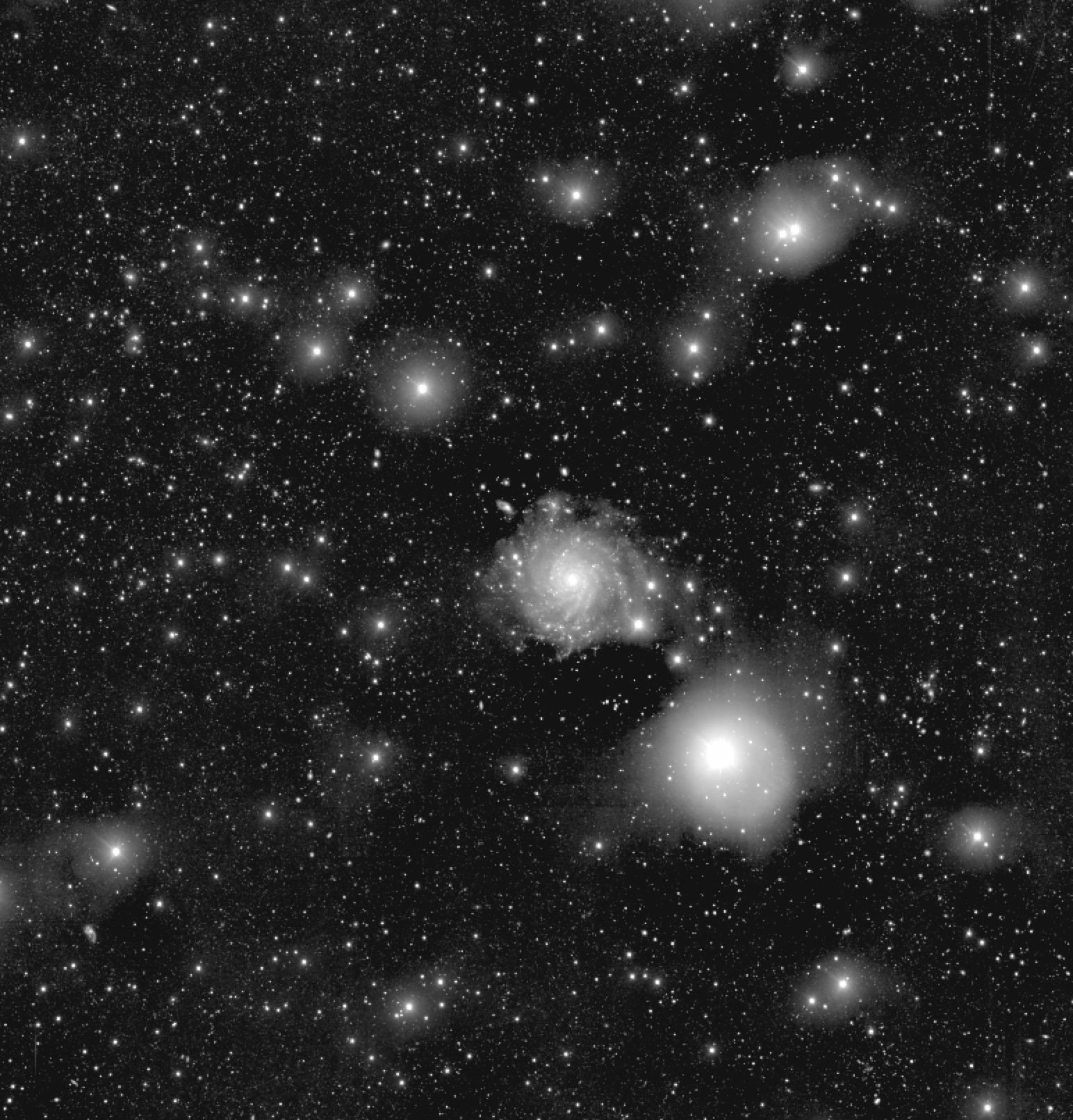}
     \caption{Left. Color composite (using $g$, $r$ and $i$ bands) VST images of  \ic5332. The image size is 28.3' $\times$ 18.6'. Right. Enlarged region of the VST FoV ($\sim$61 $\times $ 64 arcmin, i.e. $\sim$ 140 $\times $ 145 kpc) in the deep $g$-band image centred on \ic5332. North is up and East is to the left.} 
    \label{fig:composite_and_gband}
\end{figure*}

\subsection{Observations and data reduction}\label{subsec:obs_reduction}

The data utilized in this study stem from the VST-SMASH survey, an extensive imaging survey in $g$, $r$, and $i$ bands, conducted using the ESO's VST, a 2.6-meter telescope designed for wide-field optical imaging. Equipped with OmegaCAM \citep{Kuijken2002Msngr.110...15K}, which provides a 1 square degree FoV at a pixel scale of 0.21 arcsec per pixel, these observations were carried out in service mode under optimal conditions—dark skies and photometric stability with specified lunar illumination limits, benefiting from the sky conditions of Cerro Paranal (Chile). A seeing FWHM $\leq 1.4$~arcsec in all bands and a maximum airmass of 1.8 were required, along with lunar illumination below 0.3, 0.4, and 0.5 in the $g$, $r$, and $i$ bands, respectively.

The total exposure times were optimized to maximize depth, especially in the $g$ and $r$ bands, which reached 2.5 hours of integration each, while the $i$ band was observed for a total of 2 hours.
 
In the LSB regime, accurately estimating the sky background contribution is crucial. Depending on the projected radius of the galaxy, two observational strategies were employed: (a) a standard strategy, observing only the field surrounding the target (ON), or (b) an ON-OFF step-dither strategy, observing both the target field (ON) and an adjacent field (OFF). Each field consists of five pointings in the $g$ and $r$ bands, and four in the $i$ band, centered on different positions. For each pointing, a standard diagonal dither strategy was used (reproducing that tested on FDS and VEGAS data), comprising six short exposures of 300s each. Both of these approaches, optimized by their respective dithering patterns, efficiently bridge the gaps between the 32 CCDs of OmegaCAM on the VST.
For the galaxies with a projected major axis under 5 arcmin, the standard strategy was employed, and only the field around the target was observed. As documented in \citet{Capaccioli2015A&A...581A..10C} and \citet{spavone2017Galax...5...31S}, this method estimates the sky background directly from the science frame by fitting a 2D polynomial surface to the mosaic pixel values that are unaffected by celestial objects.

For more extended and bright galaxies with extensive envelopes (e.g., with major axis diameter $\gsim$ 5 arcmin), the ON-OFF step-dither technique was applied. This strategy uses the OFF field (i.e., the frame in proximity, within $\pm 1\sqd$, to the science frame but sufficiently far away from the bright, extended halo of the galaxies to avoid contamination from the faint outskirts and ICL) to independently measure the sky background, modeled as a plane, after carefully masking the foreground/background sources, and subtracts it from the whole mosaic consisting of the ON and OFF fields, in case it is also needed to manage the OFF frame. \citep[see e.g.,][]{spavone2017A&A...603A..38S,ragusa2021A&A...651A..39R,ragusa2022FrASS...952810R,ragusa2023A&A...670L..20R}.
Data reduction and calibration were processed through the Astro-WISE pipeline, developed specifically for OmegaCAM observations \citep{McFarland2013ExA....35...79M}, encompassing bias, flat-field correction, sky background subtraction, and the creation of a co-added mosaic \citep[for more details see][]{venhola17}  with a spatial scale of 0.2 arcsec per pixel.

\begin{table}
\centering
\begin{threeparttable}
\caption{Literature data for \ic5332.}
\begin{tabular}{lc}
\hline\hline
Parameter & Value \\
\hline
Target Name & \ic5332$^{(a)}$ \\
Morphological Type & SA(s)d \\
Hubble Type & 7 \\
Environment & NGC 7713 Group \\
Right Ascension (J2000) & 23:34:27.567 \\
Declination (J2000) & $-36$:06:03.728 \\
Distance (D) [Mpc] & $7.80$ \\
Linear scale [kpc/arcsec] & 0.0378 \\
$A_{26}$ [arcmin] & 8.32 \\
$A_{26}$ [kpc] & 19 \\
Ellipticity ($\epsilon$) & 0.13 \\
Inclination ($i$) [$^\circ$] & 30 \\
Total Mass ($M_\mathrm{tot} (r < A_{26})$) [\Msun] & $1.7 \times 10^{10}$ \\
\hi\ Mass ($M_{\rm H\,I}$) [\Msun] & $1.5 \times 10^{9}$ \\
Stellar mass ($\mst$) [\Msun] & $1.4 \times 10^{9}$$^{(b)}$ \\
$\rm SFR_{\rm H{\rm \alpha}}$ [$\Msun \, $yr$^{-1}$] & $0.3$$^{(c)}$ \\
$\rm SFR_{\rm FUV}$ [$\Msun \,$yr$^{-1}$] & $0.4$$^{(c)}$ \\
\hline
\end{tabular}
    \begin{tablenotes}
    \footnotesize
    \item $^{(a)}$ \cre{Most of the data are taken from \cite{Karachentsev+13}, unless otherwise stated.}
    \item $^{(b)}$ Based on BVK magnitudes from \cite{Karachentsev+13} and $M/L$-colour relations from \cite{BdeJ01}, based on a scaled Salpeter IMF. \cre{This value is much lower than the estimate of $7.6 \times 10^{9}\,\Msun$ reported by \citet{Cook+14}, obtained from the $3.6\,\mu\mathrm{m}$ emission and assuming a constant $M/L = 0.5$.}
    \item $^{(c)}$ \cre{SFRs are calculated using \ha\ and FUV data and formulae in \cite{Karachentsev+13}, and rescaled to a Chabrier IMF. These values are quite consistent with the estimates by \citet{Lee+09}, namely 0.34 and 0.5~$\Msun\,\mathrm{yr}^{-1}$ (after accounting for the IMF conversion).}\\
    \end{tablenotes}\label{tab:TARGET}
    \end{threeparttable}
\end{table}

\subsection{Surface photometry}\label{sec:SB_photometry_method}

The investigation of LSB structures linked to galaxy halos represents a major challenge in contemporary astronomy due to their extremely faint SB levels ($\mu_V \geq$ 26–27 mag/arcsec$^2$) and, in some cases, their vast spatial scale, as seen in the ICL or the large tidal tails, which can stretch over 100 kpc. These intrinsic properties make LSB features particularly difficult to analyze.

As detailed in \Sec\ref{sec:intro}, overcoming these challenges requires the use of wide-field, ultra-deep, multi-band photometry. Nevertheless, analyzing LSB structures adds a layer of complexity to interpretation.  Their similar origins through accretion events make it particularly difficult to distinguish LSB features from the original stellar envelope using photometry alone. This difficulty arises from the gradual and smooth transition between the accreted stellar halo and the surrounding LSB components, which often share comparable SB profiles and are therefore challenging to separate.

To address these complexities, the characterization of the LSB components was conducted through a robust methodology grounded in well-established surface photometry techniques, as described in our recent studies \citep[e.g.,][]{ragusa2021A&A...651A..39R,ragusa2022FrASS...952810R,ragusa2023A&A...670L..20R}. This approach allows for a detailed examination of faint galactic outskirts and potential tidal features. 

This methodology includes cropping low-S/N edges, manual masking of contaminating sources, and correction for scattered light from bright stars. 
The limiting radius ($R_{\rm lim}$) was defined where the galaxy's light merges with residual background fluctuations, using the sky RMS to estimate uncertainties. 
Isophotal fitting was finally performed in the $g$, $r$, and $i$ bands with \texttt{photutils.Ellipse}, fixing the center and deriving radial profiles of SB, ellipticity ($\epsilon$), and position angle (P.A.). 
These steps ensured a robust characterization of the LSB components. A more detailed description of the surface photometry steps is provided in \App\ref{app:SB_photometry_details}.

\section{The first VST-SMASH target: \ic5332}\label{sec:targets}

In this section, we present the target analyzed in this study, providing a concise
description of its fundamental characteristics, along with a summary of the key findings reported in the literature in \Sec\ref{subsec:IC5332}. 
In \Sec\ref{subsec:depth} we will provide a general description of the observational setup adopted for the target, the FWHM PSF in the three VST $gri$ bands, as well as the depth achieved.

\subsection{\ic5332: the literature}\label{subsec:IC5332}

\ic5332 is an almost face-on, LSB spiral galaxy located in the Sculptor constellation, at $\sim$ 7.8 Mpc distance. It belongs to the NGC~7713 group, with NGC~7713 likely acting as its primary perturber. Among the galaxies in the VST-SMASH sample, it exhibits intermediate values of stellar, total, and \hi\ mass. In particular, the stellar and \hi\ masses are both $\sim$ $10^9\,\Msun$, while the total mass is estimated to be $\sim 2 \times 10^{10}\,\Msun$. The SFR, derived from both \ha\ and FUV emission, is $\sim 0.3-0.4\, \rm \Msun \, yr^{-1}$. Its main physical properties, compiled from the literature, are summarized in \Tab\ref{tab:TARGET} and some properties are also visualized in \Fig\ref{fig:VST-SMASH_sample_properties}.

It exhibits interesting interstellar structures, particularly in the context of gas dynamics, kinematics, and star formation (SF) processes. The galaxy is part of the PHANGS-ALMA \citep{Schinnerer2019Msngr.177...36S,Leroy2021ApJS..255...19L} and PHANGS-JWST \citep{lee2023ApJ...944L..17L} programs, and its study provides valuable insights into the interrelation of gas, dust, and SF. Below, we explore key aspects of \ic5332\ in terms of its physical properties.

\begin{itemize}
    \item Star formation and stellar populations: \ic5332\ shows many regions of active SF, \cre{at a rate consistent with the SF main sequence \citep[\Fig\ref{fig:VST-SMASH_sample_properties},][]{Hunt2020,Leroy2021ApJS..255...19L,Popesso+23}.} Its SF extends to large galactocentric radii, with UV emission suggesting that the outer disk is still forming stars \citep{Hassani2024ApJS..271....2H}. However, there are no strong spiral patterns or massive star-forming complexes present. Instead, SF occurs in irregular clumps, particularly in the south-east and western regions of the galaxy. The \ha\ and FUV emission \citep{Karachentsev+13}
    supports the idea of recent SF in these regions. 
    This irregularity questions the classification of \ic5332\ as an XUV-disk galaxy \citep{Thilker2007ApJS..173..538T}.

    \item Gas properties and kinematics: \ic5332\ exhibits complex gas dynamics, with an extensive \hi\ component detected in its outer disk, as observed by Parkes and ATCA \citep{Pisano2011ApJS..197...28P}. The internal kinematics, as revealed by MeerKAT observations, suggest that \ic5332’s \hi\ disk is relatively undisturbed, with no signs of significant tidal interaction \citep{Leroy2021ApJS..255...19L}. In contrast, PHANGS-JWST data reveal a web of filamentary structure in the interstellar medium (ISM) within the galactic disk. These filaments, lacking a large-scale SF pattern, resemble cold gas filaments seen in simulations of multi-phase disk fragmentation \citep{Meidt2024ApJ...966...62M}. Such structures may result from differential rotation stretching overdensities within the disk \citep[e.g.,][]{Duarte-Cabral2017MNRAS.470.4261D}.

   \item   Metallicity and environmental impact:
    \ic 5332\ is characterized by a relatively small number of H\,II regions, which limits the spatial sampling of its gas-phase metallicity. \cre{PHANGS-MUSE data  indicate that \ic5332\  exhibits a strong gas metallicity gradient \citep{groves2023MNRAS.520.4902G,Hassani2024ApJS..271....2H}. It also present low gas metallicities, in line with the mass-metallicity relation established by \citet{lee06,Lee+09}, according to which less massive galaxies tend to be metal-poorer.} Furthermore, as discussed by \citet{leroyni2013AJ....146...19L}, low-metallicity systems like \ic 5332 are expected to show a reduced CO detectability and an increased CO-to-$H_2$ conversion factor, highlighting the connection between metallicity and the molecular gas content.
    Though part of the NGC 7713 galaxy group, \ic5332\ does not show signs of active interaction with nearby galaxies. However, its location within this group and irregular SF patterns suggest the possibility of past mergers or interactions that may have influenced its outer disk and \hi\ structure \citep{Pisano2011ApJS..197...28P}.
\end{itemize}

Therefore, \ic5332\ presents an intriguing case of a low-mass, star-forming galaxy with a unique interplay between its gas, dust, and star-forming components. The lack of strong spiral arms and ordered star-forming complexes, combined with a complex filamentary gas structure, sets \ic5332\ apart from other galaxies in the PHANGS-JWST sample. While the galaxy's metallicity gradient and PAH properties are consistent with its low SFR, its outer disk may have undergone episodes of interaction, contributing to its current state. Further study of its star-formation history (SFH) will help clarify its evolution and the role of the environment in shaping its properties.

\subsection{Observational set-up, PSF and survey depth}\label{subsec:depth}

\ic5332\ was observed between May and December 2023 as part of the VST-SMASH runs 110.25AA.001 and 112.266Z.001, under optimal seeing conditions. Due to the galaxy’s large angular size, the ON–OFF strategy was adopted. The OFF frame was used not only to estimate the sky background around \ic5332\ but also for the other adjacent field containing NGC 7713, resulting in a mosaic covering approximately $3 \, \sqd$. In this paper, however, we focus solely on the surface photometry analysis of \ic5332, and refer the reader to a forthcoming paper for the analysis of the remaining field.

The colour-composite image, created by combining the three bands, \cre{is shown in the left panel of \Fig\ref{fig:composite_and_gband}, while the $g$-band image covering an area of approximately 1 square degree is presented in the right panel of \Fig\ref{fig:composite_and_gband}.} This illustrates how a single VST FoV can encompass a significant portion of the DM halos of galaxies targeted by the VST-SMASH survey. In the case of \ic5332, the image allows us to trace its outskirts out to $\sim 70\,\rm kpc$. Furthermore, by leveraging the larger mosaic that includes NGC\,7713, we can potentially probe the DM halo of \ic5332\ out to a distance of $\sim 350\, \rm kpc$.

We calculate the PSF FWHM by fitting a Moffat function to the stars with the highest S/N, excluding saturated stars, and restricting the analysis to stars in the vicinity of \ic5332, after masking a circular region of 150 arcsec radius centered on the galaxy. From this analysis, we determine median FWHM values of $\sim 0.9$, $\sim 1.1$, and $\sim 0.8$ arcsec in the $g$, $r$, and $i$ bands, respectively. These values meet the intended observational requirements to conduct studies at both high spatial resolution and great depth, which are essential to accurately trace and characterize LSB features in the outskirts of galaxies \citep[e.g.,][]{ragusa2021A&A...651A..39R}.

Finally, calculating the limiting magnitudes in each band is crucial for verifying our ability to detect and characterize faint, diffuse structures. In the LSB regime, achieving reliable measurements of these limits ensures that we can identify and analyze extended features, such as stellar halos, tidal debris, and other faint structures, that are often concealed within background noise. We use the methods outlined in \cite{hunt2025A&A...697A...9H} and summarized in \App\ref{app:limiting_SB_calculations}, to calculate the formal limiting SB, \mulim, within an area of 100 arcsec$^2$ at the $1\sigma$ level. We find $1\sigma$ depths of 30.7, 30.5, and 29.5 mag arcsec$^{-2}$ in the $g$, $r$, and $i$ bands, respectively, which match the planned depths of VST–SMASH \citep{Tortora+24_VST-SMASH}, typically required for LSB analyses \citep[e.g.,][]{martinez10,vanDokkum+14}. These correspond to $1\sigma$ noise levels for individual pixels of
26.45, 26.15, and 25.25\,mag\,arcsec$^{-2}$ for $g$, $r$, and $i$ bands, respectively.

\begin{figure}
\centering
\includegraphics[width=1\linewidth]{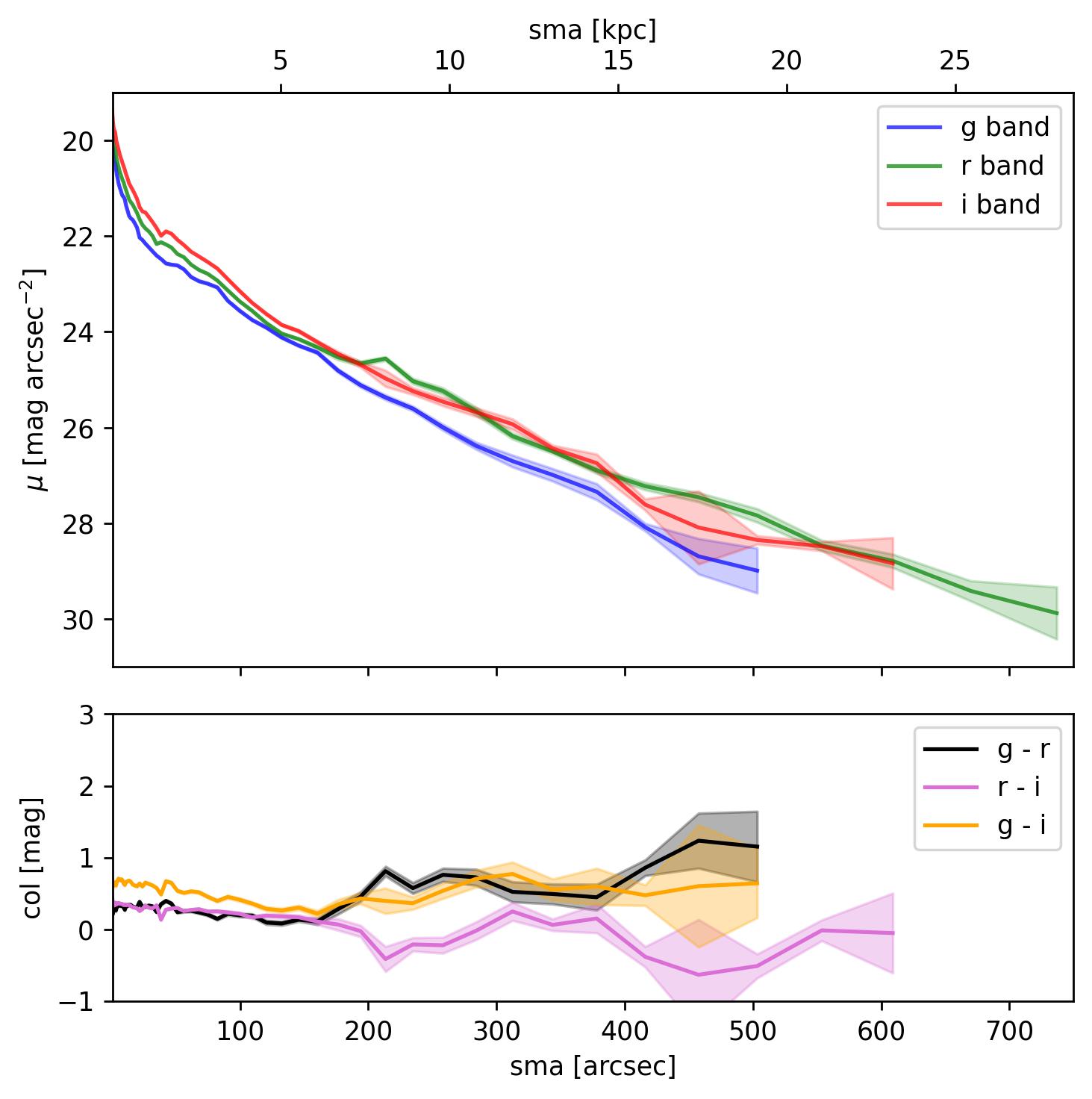}
\caption{Deep photometry of \ic5332. Azimuthally-averaged surface brightness profiles of \ic5332\ in VST $g$ (blue), $r$ (green)
and $i$ (red) bands. The $g-r$ (black),  $r-i$ (magenta) \cre{and $g-i$ (orange)} color profiles of \ic5332\ are also shown in the bottom region of the panel.}
    \label{fig:IC5332_profiles}
\end{figure}

\section{\ic5332\ deep photometry }\label{sec:SB_and_colour_profiles}

\ic5332\ is face-on galaxy, often noted for its flocculent spiral structure and low inclination, making it ideal for disc structure studies.
In this section, we present the results of the deep surface photometry analysis of \ic5332\  in the $g$, $r$, and $i$ bands, including the two-component Sérsic decomposition, as well as an investigation of its faint outer regions.

\subsection{Deep and extended surface brightness and color profiles}\label{sec:SB_5332}

The averaged radial SB profiles in the $g$, $r$, and $i$ bands, obtained
following the detailed procedure described in App. \ref{app:SB_photometry_details}, using the
initial geometric parameters from \citet{Jarrett2003AJ....125..525J}, are shown
in the top panel of Fig. \ref{fig:IC5332_profiles}. The profiles extend out to 503, 737,
and 609 arcsec in the $g$, $r$, and $i$ bands respectively, corresponding
to galactocentric distances of approximately 19, 28, and 23 kpc.
The faintest SB levels reached are 29.1, 29.9, and 28.9 mag
arcsec$^{-2}$ in the $g$, $r$, and $i$ bands,respectively. These SB limits are slightly brighter than the ones we infer over 100 arcsec$^2$ regions, because they are averaged over radial bins with areas smaller than this. Specifically, assuming a radial extent of 737 arcsec (609 arcsec) for $r$ ($i$), respectively, we would expect differences of SB limits between the profiles and those given in Sect. \ref{subsec:depth} of 0.54 mag arcsec$^{-2}$ in $r$, and 0.65 mag arcsec$^{-2}$ in $i$. These expected differences are within 0.05 mag arcsec$^{-2}$ of the two sets of SB limits given here.

Although our formal depth estimates (calculated in Sect.~3.2) suggest
comparable sensitivities in the $g$ and $r$ bands, the automated
data–reduction procedure affects the $g$-band image, producing
a residual oversubtraction in the galaxy’s outer regions. We applied
extensive masking to the areas most affected by this issue, but minor
residuals may still be present. This problem is likely related to the
background estimation stage, during which the Astro-WISE automatic
pipeline may not have properly masked nearby saturated bright stars,
leading to an overestimation of the sky level to be subtracted. As a
consequence, the oversubtraction in the $g$ band prevents us from
tracing its surface-brightness profile down to the same SB levels
reached in the $r$-band image. However, this limitation does not impact
our conclusions, which rely mainly on colour profiles that are
constrained by the shallower band.

The azimuthally averaged SB profiles, after explicitly correcting
for large-scale residual background fluctuations, display a smooth
decline with radius, which approximates an exponential trend at
intermediate-to-large radii (see next sections about structural
analysis), as typically observed in disk-dominated galaxies.
The lower panel of Fig.~3 presents the radial color profiles in
$g-r$, $r-i$, and $g-i$. Also in this case, we allowed the ellipse
geometry to vary freely in each band, in order to better trace possible
physical variations within the galaxy, since the main goal of our work
is to highlight potential asymmetries related to accretion events.
However, we emphasize that this choice, which appeared to be the most
consistent with our scientific aims, does not significantly affect the
results, as can be seen in Fig.~5 (see Sect.~4.4), where the radial
trends of $\epsilon$ and PA show no substantial variation from band to
band.

Within the inner $\sim 40$ arcsec ($\sim 1.5$ kpc), the colors are
remarkably flat, indicating a relatively homogeneous stellar
population in the central regions. A trend toward bluer colors is
observed out to $\sim 170$ arcsec ($\sim 6.5$ kpc), suggesting the
presence of a radial stellar population gradient, with younger and/or
lower-metallicity stars populating the outer disk
(e.g., \citealt{Tortora+10CG}).

Beyond this radius, the color gradients exhibit an inversion: the
$g-r$ and $g-i$ colors become redder, while the $r-i$ trends toward
bluer values, although with larger uncertainties. The $g-r$ and $g-i$
behavior might indicate the emergence of distinct stellar components
in the outskirts, such as faint tidal debris, stellar halos, or LSB
streams (see Sect.~5), where photometric uncertainties increase
due to the dominance of background fluctuations. In this regime,
redder $g-r$ and $g-i$ colors can be interpreted as the contribution
from older and/or more metal-rich stellar populations, while the
decline in $r-i$ color may require particular caution.

Interestingly, the colour inversion observed in the outer regions—
where $g-r$ becomes redder and $r-i$ bluer—and, in particular, the
significant bump at $\sim 220$ arcsec in the $r$ band, may indicate
enhanced $r$-band emission. Therefore, a contribution from diffuse
H$\alpha$+[N\,\textsc{ii}] emission cannot be excluded—particularly in
the presence of extraplanar ionized gas—even if it is difficult to
distinguish in an almost face-on galaxy like IC~5332. Moreover, the
reddening of the $g-r$ and $g-i$ colour gradients starts at about
170 arcsec, which—as discussed in Sect.\ref{sec:stream}—corresponds to the radial
distance where the two stellar streams detected in the outskirts of
IC~5332 begin, possibly reflecting the emergence of a distinct stellar
population associated with accreted material. 

To assess whether the colour behaviour in the radial range 
$170 \lesssim R \lesssim 400$ arcsec could be affected by scattered 
light from bright foreground stars, we performed an angular sector 
analysis of the surface brightness and colour profiles. The galaxy was divided into three independent sectors, one of which (-60 \textdegree, 60\textdegree), anticlockwise with respect to the horizontal, completely includes the region where the bright stars are located. The resulting 
colour profiles are shown in \Fig \ref{fig:sectors}. We find that, within the radial 
range of interest, the sector encompassing the region potentially affected by scattered stellar light shows no significant deviation with respect to the others. This consistency indicates that the measured colour properties in the 
outer regions are not significantly affected by stellar contamination. 
A detailed description of this test is provided in  App. \ref{app:quadrants}.

\subsection{\cre{Structural parameters from the growth curve and colour gradients}}\label{subsec:growth curve}

The growth curve analysis provides, for such extended galaxies, a model-independent approach for estimating the galaxy effective radii ($R_e$). To obtain the growth curve, we integrated the SB profiles up to the radial extension of each band, reported in \Sec\ref{sec:SB_5332} and in \Fig\ref{fig:IC5332_profiles}\footnote{Although some residual oversubtraction is visible in the $g$ band, the growth curve flattens out toward a plateau, confirming that the results are only marginally affected by this issue.}. Once the \Re\ values were derived for each band, we computed the integrated magnitudes and colours of \ic5332\ within this radius. 
The resulting effective radii are $R_{e,g}$ = 108 arcsec ($4.1$ kpc), $R_{e,r}$ = 127 arcsec  ($4.8$ kpc), and $R_{e,i}$ = 104 arcsec  ($3.9$ kpc). 
They indicate that in the $r$ band, the galaxy is less concentrated than in the $g$ and $i$ bands. This could also be an effect of the enhanced emission in the $r$ band seen both in the bump at 220 arcsec and in the outer regions. Therefore, these values show an increase from $g$ to $r$ band and then a decrease in $i$ band, indicating a non-monotonic trend with wavelength, contrasting with the average trends found in observations \citep{Vulcani+14} and simulations \citep{Baes+24_II} of late-type galaxies. This suggests a more intricate interplay among stellar population parameters, dust, and the effect of emission lines. When compared with a population of local late-type galaxies selected from the NYU-VAGC catalog \citep{Blanton+05_NYU}, according to the selection criteria on Sérsic index and concentration adopted in \cite{Tortora+10CG} the ratios of the effective radii in the different bands appear to be broadly consistent with the bulk of the local galaxy population. To be more specific, we find $R_{\rm e,g}/R_{\rm e,r} = 0.86$ and $R_{\rm e,g}/R_{\rm e,i} = 1.04$, where the former is smaller than the median value reported in the NYU-VAGC catalog at similar stellar masses, while the latter is consistent with the corresponding median (i.e., $R_{\rm e,g}/R_{\rm e,r} = 1.02 \pm 0.07$ and $R_{\rm e,g}/R_{\rm e,i} = 1.03 \pm 0.10$).

Because the ratios of the $R_{e}$ provide only partial information, we follow \citet{Tortora+10CG} and calculate the colour gradient between $0.1 \times R_{\rm e,r}$ and $R_{\rm e,r}$. Studying the correlations between colour and stellar population gradients and galaxy mass is known to provide deeper constraints on galaxy formation and evolutionary processes \citep[e.g.][]{Tortora+10CG,Tortora+11CGsim,Tortora+11MtoLgrad,Tortora+13_CG_SIM,goddard2017MNRAS.465..688G,liao2023MNRAS.518.3999L}. We adopt the $g-i$ colour for consistency with \citet{Tortora+10CG} and to exclude the $r$ band, which can be affected by \ha-\ntwo\ emission, as discussed previously. The colour profile is fitted with the relation $g-i = c + m \, \times \, \log (R/R_{\rm e,r})$, where $m$ is the colour gradient. We find a gradient of $m = -0.32$ dex, which is steeper than the median value for LTGs of similar stellar mass ($\sim 1$–$3 \times 10^{9} \, \Msun$), reported by \citet{Tortora+10CG} as $\sim -0.15$ dex. 

To interpret these colour gradients, we note that \citet{Groves+23_PHANGS-MUSE} report an internal dust extinction for \ic5332\ decreasing from $A_{V} = 0.40$ mag at the centre to $A_{V} = 0.25$ mag at one \Re. Assuming a linear variation of extinction between these radii, we derived the corresponding corrections in the $g$ and $i$ bands and applied them to the colour profiles. After removing the dust contribution, the resulting gradient is $m = -0.24$ dex, which is consistent with being driven by stellar metallicity variations \citep{Tortora+10CG}. A direct comparison with the results of \citet{Tortora+10CG} is not possible, since an analogous dust correction cannot be applied to their sample. Nevertheless, a metallicity-driven gradient, with the metallicity radially decreasing, is in line with the steep gas-phase metallicity gradients of \ic5332\ measured in \citet{Groves+23_PHANGS-MUSE}, which are among the steepest in the PHANGS sample. 

Steep metallicity gradients in massive galaxies can be largely explained within the framework of the monolithic collapse scenario, which is further enhanced by the contributions of stellar and supernova-driven outflows. They are expected when stars form during strong dissipative (monolithic) collapses in the deep potential wells of galaxy cores, where the gas is more efficiently retained. This results in more efficent SF and chemical enrichment in the central regions compared to the outskirts, thereby producing negative metallicity gradients \citep[e.g.,][]{Larson74,Kobayashi04}. In addition, the delayed onset of supernova-driven winds, which provide a further supply of metals to the central regions, can act to reinforce the steepness of these gradients \citep[e.g.,][]{Pipino+08}. Since the depth of the gravitational potential regulates these processes and thus correlates with galaxy mass, massive galaxies are expected to host steeper metallicity gradients than lower-mass ones. The sharp colour (and presumably metallicity) gradients observed in \ic5332\ can be naturally explained within this scenario.

The total magnitudes measured out to the outermost radius reached by the photometric profiles 
are: $m_g$ = 10.68 mag, $m_r$ = 10.29 mag, and $m_i$ = 10.19 mag. As expected, these values show a progressive brightening from the $g$ to the $i$ band. 
We also computed the integrated magnitudes within the $r$-band $R_e$. These values are $m_g$ = 11.29 mag, $m_r$ = 11.04 mag, and $m_i$ = 10.78 mag, which also show a consistent brightening toward redder bands.  

We also estimated the isophotal (or Holmberg) radius corresponding to an SB level of 26.5 mag arcsec$^{-2}$, obtaining values of 294, 344, and 350 arcsec in $g$, $r$, and $i$, respectively, that are within the maximum radial extent reached by the SB profiles (see \Sec\ref{sec:SB_5332}). Comparing with the SB profiles, this confirms that light is detected well beyond the 26.5 mag arcsec$^{-2}$ isophote, extending to radii of about twice the Holmberg radius.
This highlights the importance of deep photometry in capturing the extended LSB emission in spiral galaxies, which may significantly affect the inferred stellar luminosity and mass budget (see \Sec\ref{sec:mass}), especially when a substantial fraction of the light lies beyond the $R_e$ and well past conventional isophotal radii.
From these values, we derived both the total integrated colors and those computed within the $R_e$ (corrected only for foreground extinction). The total colours are $g - r = 0.39$ mag and $r - i = 0.10$ mag, indicating an overall blue stellar population. 

The colors measured within the $R_e$ are $(g - r)_{R_e}  = 0.25$ mag and $(r - i)_{R_e} = 0.26$ mag, showing that the inner regions are slightly bluer in $g - r$ but redder in $r - i$ compared to the global values. These differences are clearly driven by the opposite behaviour of $g-r$ and $r-i$ in the outer regions.
The total integrated colors derived from the growth curve method are fully consistent with those expected for less massive spiral galaxies. These values fall within the typical ranges observed for Sc–Sd galaxies, which generally show $g-r$ in a range 0.3-0.5 mag and $r-i$ between 0.1 and 0.3 mag \citep[e.g.][]{Strateva+01, baldry2004ApJ...600..681B}.  Such colors reflect the composite nature of the stellar populations in these systems: the relatively blue $g-r$ color indicates ongoing SF and the presence of young stellar populations in the disk, while the modest $r-i$ color suggests a moderate contribution from more evolved stars, particularly in the inner regions.
Overall, \Re, magnitudes, and colors show systematic trends with wavelength, consistent with a stellar population increasingly dominated by old stars towards redder bands.

\subsection{Sérsic decompositon}\label{subsec:decomp}

In this section, we model the SB profile of the galaxy in the $g$, $r$, and $i$ bands. 
The analysis is performed on azimuthally averaged surface-brightness profiles extracted using isophotes whose ellipticity and position angle are allowed to vary with radius, in order to accurately trace the intrinsic geometry of the galaxy and possible physical asymmetries. At each radius, the azimuthal averaging provides a robust estimate of the mean light distribution, which is well-suited for one-dimensional structural modeling.

As shown in \App\ref{app:sersic1comp}, a single-component Sérsic model fails to provide a satisfactory fit to the galaxy’s outer regions, which require an additional exponential component, even though these outer parts contribute only marginally to the total stellar mass. Therefore, because a single Sérsic component does not adequately reproduce the full SB profile in any of the three bands, particularly in the outer regions, we introduced a second component. 

\begin{figure}
\centering
\includegraphics[width=1 \linewidth]{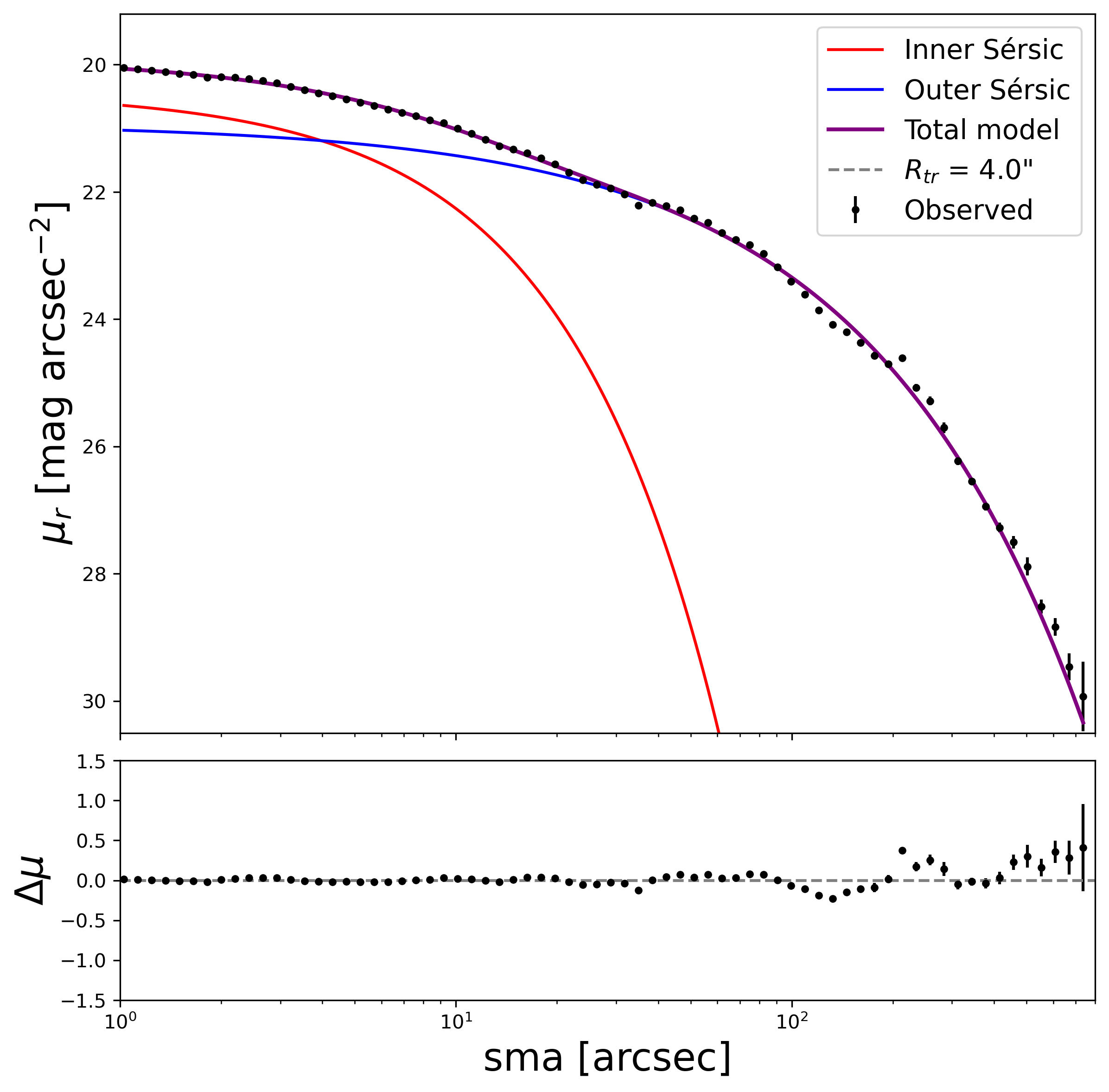}
\caption{Sérsic decomposition of \ic5332. 
Top panel: Two-component model of the azimuthally averaged surface brightness profiles of \ic5332\ in the $r$ band. The red line indicates the inner Sérsic component, while the blue one corresponds to the outer regions. The purple line represents the sum of both components as the best-fit for the observed light profile. 
The black vertical dashed line shows the estimated value for R$_{tr}$.
Bottom panel: The bottom panels show the residuals of the fit, $\Delta\mu = \mu_{obs} - \mu_{model}$.}
    \label{fig:IC5332_fits}
\end{figure}

The Sérsic decomposition is intentionally performed in one dimension, as the primary goal of this analysis is to characterize the radial structural components of IC~5332 as a function of galactocentric distance, rather than to perform a full two-dimensional morphological decomposition of the galaxy.
The SB profiles of \ic5332\ in the $g$, $r$, and $i$ bands were therefore modeled with a 1D two-component Sérsic decomposition using our Python code. The results are presented in \Fig\ref{fig:IC5332_fits}. 
Parameter uncertainties were estimated as the square roots of the diagonal elements of the covariance matrix provided by the fitting algorithm curve$_{\rm fit}$ within the Python library scipy.optimize. This matrix is computed from the inverse of the Hessian of the $\chi^2$ function, and the resulting values represent the 1$\sigma$ confidence intervals under the assumption of Gaussian, independent errors.
The fitting is performed using inverse-variance weighting, where the uncertainties associated with each radial bin include both photon noise and large-scale background fluctuations, as estimated following the procedure described in App. \ref{app:SB_photometry_details}.
The goodness of fit is quantified using the reduced $\chi^2$ and the sum of the absolute residuals, allowing a direct comparison between single- and two-component Sérsic models.
In all bands, the two-component Sérsic model provides a statistically better description of the data, yielding lower reduced $\chi^2$ values and the sum of the absolute residuals than the single-component fit (see App. \ref{app:sersic2comp_app}  and \ref{app:sersic1comp} for details).
Each profile was fitted with an inner (red) and an outer (blue) Sérsic component, highlighting the radial structural variation as a function of wavelength. The inner component, typically associated with a central stellar concentration or an inner disc, shows a clear trend: the effective SB becomes progressively brighter from $\mu_{e1,g} = 22.60 \pm 0.11$ to $\mu_{e1,r} = 22.37 \pm 0.06$ and $\mu_{e1,i} = 21.86 \pm 0.07$ mag arcsec$^{-2}$. Simultaneously, the effective radius decreases from $R_{e1,g} = 12.87 \pm 0.6$ arcsec to $R_{e1,r} = 10.65 \pm 0.5$ arcsec and $R_{e1,i} = 10.35 \pm 0.4$ arcsec, while the Sérsic index $n_1$ evolves from $n_{1,g} = 1.17 \pm 0.06$ to $n_{1,r} = 1.06 \pm 0.03$ and $n_{1,i} = 0.99 \pm 0.05$. This behavior suggests that the inner structure becomes more compact and dominant at longer wavelengths, consistent with a central concentration of older stellar populations. The outer Sérsic component, which likely traces the extended disk, also exhibits systematic variations. The effective SB remains nearly constant in $g$ and $r$ ($\mu_{e2,g} = 23.80 \pm 0.04$ and $\mu_{e2,r} = 23.77 \pm 0.06$ mag arcsec$^{-2}$) and becomes brighter in $i$ ($\mu_{e2,i} = 23.30 \pm 0.06$ mag arcsec$^{-2}$). The effective radius shows some fluctuations, with $R_{e2,g} = 114.85 \pm 3$ arcsec, $R_{e2,r} = 126.59 \pm 4$ arcsec, and $R_{e2,i} = 110.34 \pm 4$ arcsec. However, the Sérsic index shows a marked increase from $n_{2,g} = 1.06 \pm 0.06$ to $n_{2,r} = 1.47 \pm 0.1$ and $n_{2,i} = 1.37 \pm 0.1$. This suggests that the outer profile becomes more centrally concentrated at longer wavelengths, reflecting a growing contribution from older stellar populations dominating the light in the $r$ and $i$ bands. The transition radius $R_{tr}$, defined as the point where the two components have equal SB, also varies significantly: it decreases from $R_{tr,g} = 9.11$ arcsec to $R_{tr,r} = 3.98$ arcsec, and reaches $R_{tr,i} = 5.20$ arcsec. The reduced $R_{tr}$ in the redder bands compared to the $g$ band indicates that the outer disk component begins to dominate the light distribution much closer to the center at longer wavelengths, while the larger $R_{tr,g}$ reflects the wider dominance of the younger stellar population in the central regions.

Finally, the best-fit model (violet curve) in each band successfully reproduces the observed light distribution across the full radial range, confirming the presence of two structural components: the inner component, associated with a more compact central stellar structure, or inner disc, formed through secular processes such as bar-driven inflows or early gas collapse, in conjunction with supernovae feedback (driving the negative colour gradients discussed in \Sec\ref{subsec:growth curve}) and an outer Sérsic component, which likely traces an extended stellar disc that could be a result of radial migration, accretion of material, or minor interactions. 
Altogether, these findings reinforce the importance of multi-band surface photometry in tracing stellar population gradients and disentangling the structural components of disk galaxies, especially considering the observed dependence of the galaxy’s structural parameters on the wavelength, pointing out that the observed light distribution is not solely shaped by structural morphology but is strongly modulated by the age and metallicity distributions of the stellar content.

\subsection{Tracing the morphology}\label{ell_pa}

\begin{figure}
\centering
\includegraphics[width=8.9cm]{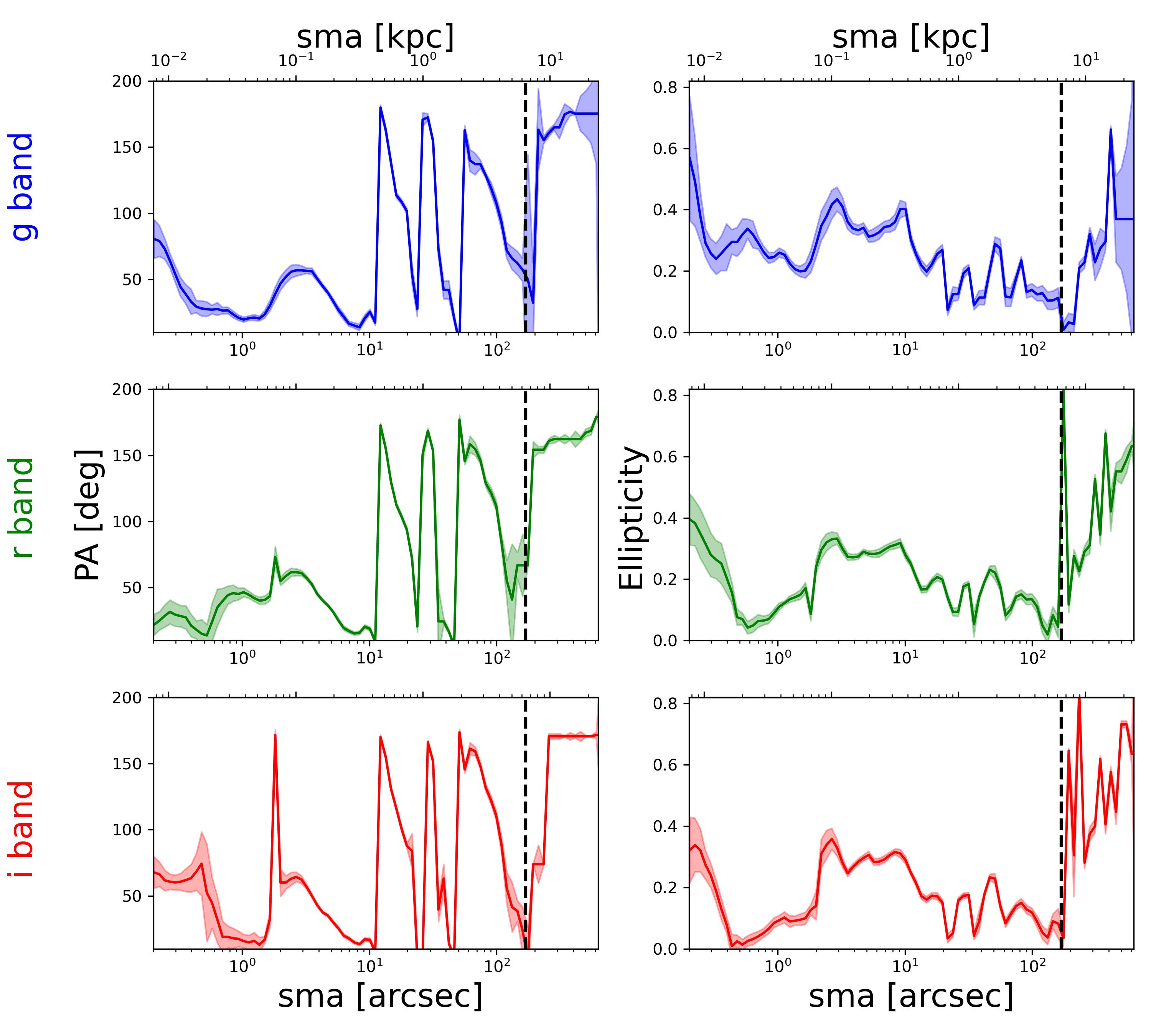}
\caption{Radial profiles of position angle (left column) and ellipticity, \cre{defined as $\epsilon = 1 - (b/a)$ (right column)} as a function of semimajor axis (in arcsec) for the galaxy \ic5332, in the $g$ (top), $r$ (middle), and $i$ (bottom) bands. Shaded regions indicate the uncertainties. The vertical dashed line marks a transition where stellar streams or asymmetric low surface brightness structures begin to dominate the isophotal shape and orientation.}\label{fig:ELL_PA}
\end{figure}

The PA and $\epsilon$ radial profiles, measured in the $g$ (blue), $r$ (green), and $i$ (red) bands, are shown in \Fig\ref{fig:ELL_PA}, in the left and right panels, respectively. In all three bands, the inner regions display a relatively smooth behavior for the PA and $\epsilon$ values, consistent with a nearly face-on spiral disk morphology.
At intermediate radii (from $\sim$ 12 to $\sim$ 170 arsec), the PA profiles exhibit a series of abrupt oscillations, evident in all three bands. These fluctuations are likely associated with the presence of spiral arms, whose asymmetric and winding structures locally distort the isophotes, leading to local twists in the isophotes that propagate into the PA measurements. The $\epsilon$ profiles in this radial range remain moderately low and stable, showing a gentle decline with radius, further confirming that the main disk is nearly circular in projection. In particular, considering radii outside the central bulge, we find that the $\epsilon$ are mostly below 0.2, corresponding to a circular disk with an inclination of $\lesssim 35^\circ$, consistent with the values reported in Table~\ref{tab:TARGET} and in the literature. 
However, at larger radii, particularly beyond $\sim$ 170 arcsec, both the PA and $\epsilon$ undergo significant changes. The PA tends to stabilize at high values ($\sim$ 170–180 $\deg$), forming a plateau, while the $\epsilon$ starts to rise, reaching values as high as at least $\epsilon$ $\sim$ 0.7 in all bands. This behavior marks a transition in the photometric morphology, where the isophotes are no longer dominated by the disk but instead by more asymmetric and elongated outer features, such as stellar streams or LSB substructures, which could alter both the orientation and the shape of the isophotes.
This coherent change in both PA and $\epsilon$ across the three bands, despite slight differences in the amplitude of fluctuations, strongly supports the scenario in which the outer isophotes are dominated by tidal material, likely of external origin. Such features are characteristic of ongoing or past accretion events, and their photometric imprint is visible both in the radial profiles and in the morphological deviations from the inner, symmetric, star-forming disk-dominated structure.

\begin{table*}[ht]
\centering
\begin{threeparttable}
\caption{\cre{Summary of the coefficients for the adopted $M/L_{r} - (g-r)$ and $M/L_{r} - (g-i)$ relations.}}
\begin{tabular}{lcccccl}
\hline
\hline
\multirow{2}{*}{Reference} & \multicolumn{2}{c}{$g-r$} & \multicolumn{2}{c}{$g-i$} & IMF / Model details \\
\cline{2-5}
 & $a_{r,g-r}$ & $b_{r,g-r}$ & $a_{r,g-i}$ & $b_{r,g-i}$ &  \\
\hline
Z09 & $-0.84$ & $1.654$ & $-0.977$ & $1.157$ & Chabrier IMF, Marigo TP-AGB prescription \\[3pt]
RC15 (BC03) & $-0.792$ & $1.629$ & $-0.861$ & $1.110$ & Chabrier IMF, BC03 with Girardi TP-AGB \\[3pt]
RC15 (FSPS) & $-0.647$ & $1.497$ & $-0.644$ & $0.973$ & Chabrier IMF, FSPS with Marigo TP-AGB \\[3pt]
IP13 (exp.\ SFH) & $-0.593$ & $1.373$ & $-0.652$ & $1.005$ & Kroupa IMF$^{(1)}$, exponential SFH models \\[3pt]
IP13 (disc templates) & $-0.663$ & $1.530$ & $-0.702$ & $1.098$ & Kroupa IMF$^{(1)}$, disc galaxy templates \\
\hline
\end{tabular}\label{tab:ML_relations}
\begin{tablenotes}
        \footnotesize
        \item $^{(1)}$ \cre{These models assume a Kroupa IMF from \citet{Rebolo+98}. To convert these to a Chabrier-equivalent IMF, we subtract 0.05 dex from the resulting $\log M/L_r$ values \citep{Tortora+09AGN}.}
    \end{tablenotes}
\end{threeparttable}
\end{table*}

\subsection{Mass and mass density profiles}\label{sec:mass}

From the available optical VST-SMASH photometry, it is not possible to constrain the stellar population parameters, but simple prescriptions can be used to estimate the stellar mass profile. We derive the stellar mass surface density and the corresponding integrated mass profile by converting the observed $g - r$ and $g-i$ colour profiles into a stellar $M/L$ using empirical $M/L$–colour relations from the literature, which are expected to depend only mildly on stellar population parameters and dust extinction. 

To estimate the $r$-band $M/L_r$ in each radial bin we adopt five different $M/L_r$–colour relations, all expressed in the form:

\begin{equation}
\log M/L_r = a_{r,col} + b_{r,col} \times \rm col,
\end{equation}

We use five different $M/L_r$–color relations: the one calibrated by \citet[][hereafter Z09]{Zibetti+09}, the two relations from \citet[][RC15]{Roediger_Courteau15}, based on different stellar population synthesis models, and the two from \citet[][IP13]{Into_Portinari2013}.
The corresponding coefficients, together with details on the adopted IMF and model prescriptions, are reported in \Tab\ref{tab:ML_relations}. To exploit the information from both color indices while keeping the analysis as simple as possible, we adopt the mean $M/L_r$ value obtained by averaging the $M/L_r$–$(g-r)$ and $M/L_r$–$(g-i)$ estimates.

In each radial bin, we calculate the absolute $r$-band luminosity density by combining the galaxy's distance modulus, the observed SB in the $r$-band, and the solar $r$-band absolute magnitude $M_{\odot,r} = 4.65$ mag. The stellar mass surface density is then obtained by multiplying the luminosity density by the corresponding $M/L_r$ value. Integration of the mass density profile yields the projected cumulative mass profile.

\begin{figure}
\centering
\includegraphics[width=0.85\linewidth]{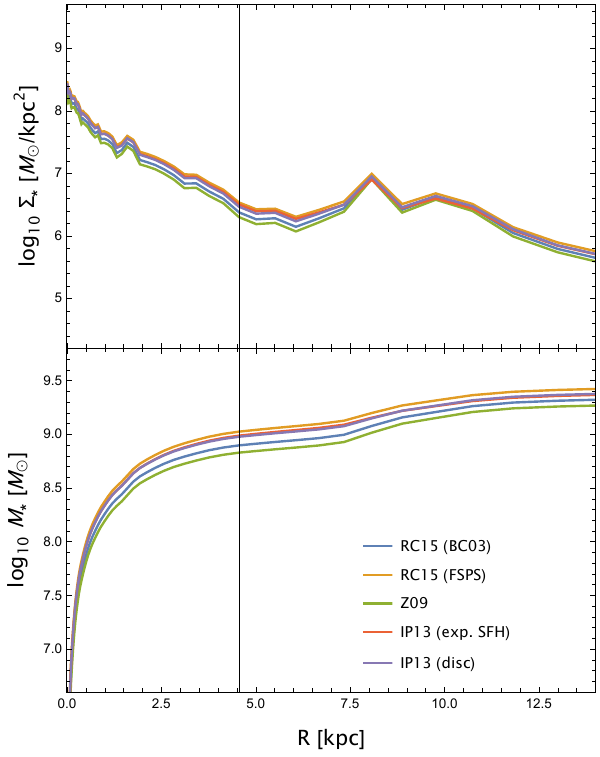}
\caption{Surface mass density (top panel) and cumulative stellar mass (bottom panel) as a function of the radius in kpc. \cre{The different colours correspond to different stellar $M/L$--color relations from the following papers: \citet{Roediger_Courteau15} (using the BC03 and FSPS models), \citet{Zibetti+09}, and \citet{Into_Portinari2013} for the exponential SFH and disc models. The $r$-band effective radius from the growth curve is shown as a vertical black line.}}\label{fig:mass_profiles}
\end{figure}

Figure \ref{fig:mass_profiles} shows the resulting mass density and integrated mass profiles for all five $M/L_r$–colour prescriptions. These models span a range of assumptions regarding stellar population synthesis. The differences among them result in an average offset between the extreme models of $\sim 0.18$ dex in $\log M/L$, corresponding to a scatter of approximately $0.09$ dex. The cumulative total stellar mass of \ic5332\ is in the range of $\log \mst/\Msun \sim 9.42-9.54$.
Following the $g-r$ and $g-i$ colour profiles, the $M/L$ remains nearly constant out to $\sim 150$ arcsec (i.e. $\sim 5.63$ kpc), beyond which it increases, reaching a peak around $R \sim 200$ arcsec (i.e. $\sim 7.5$ kpc). This complex radial trend in $M/L$ significantly affects the mass density profile, which begins to rise beyond $\sim 150$ arcsec (i.e. $\sim 6$ kpc), before reaching a maximum and then declining again. As a result, the cumulative stellar mass profile exhibits a flattening at $R \sim 6$ kpc, reaching a value of the integrated stellar mass (in M$_{\star}$) in the range $8.90-9.10$ dex, followed by a secondary rise driven by the $M/L$ bump, and finally tends to saturate at $R \gtrsim 12$ kpc, up to $\log \mst/\Msun \sim 9.3-9.4$ dex. This complex behaviour coincides with the region where a distinct stream-like feature is observed. We will explore the nature of this structure in \Sec\ref{sec:stream}.
Finally, we also derive an independent stellar mass estimate based on integrated colours up to the outermost radii. We find total masses that are consistent with those derived by integrating the mass profile. These estimates are slightly larger than the value reported in \Tab\ref{tab:TARGET}, which is based on a scaled Salpeter IMF.\footnote{The scaled Salpeter IMF of \cite{BdeJ01} yields only slightly larger values than a Kroupa IMF.} Our mass estimates are also smaller than the value of $7.6 \times 10^{9}\,\Msun$ reported by \citet{Cook+14}, obtained from the $3.6\,\mu\mathrm{m}$ emission and assuming $M/L_{3.6\,\mu\mathrm{m}} = 0.5$. This comparison confirms that such a mass estimate could be overestimated, due to their assumption of a constant $M/L$ across their entire galaxy population \citep{hunt19}.

\begin{figure*} 
\centering
{\includegraphics[width=0.47\linewidth]{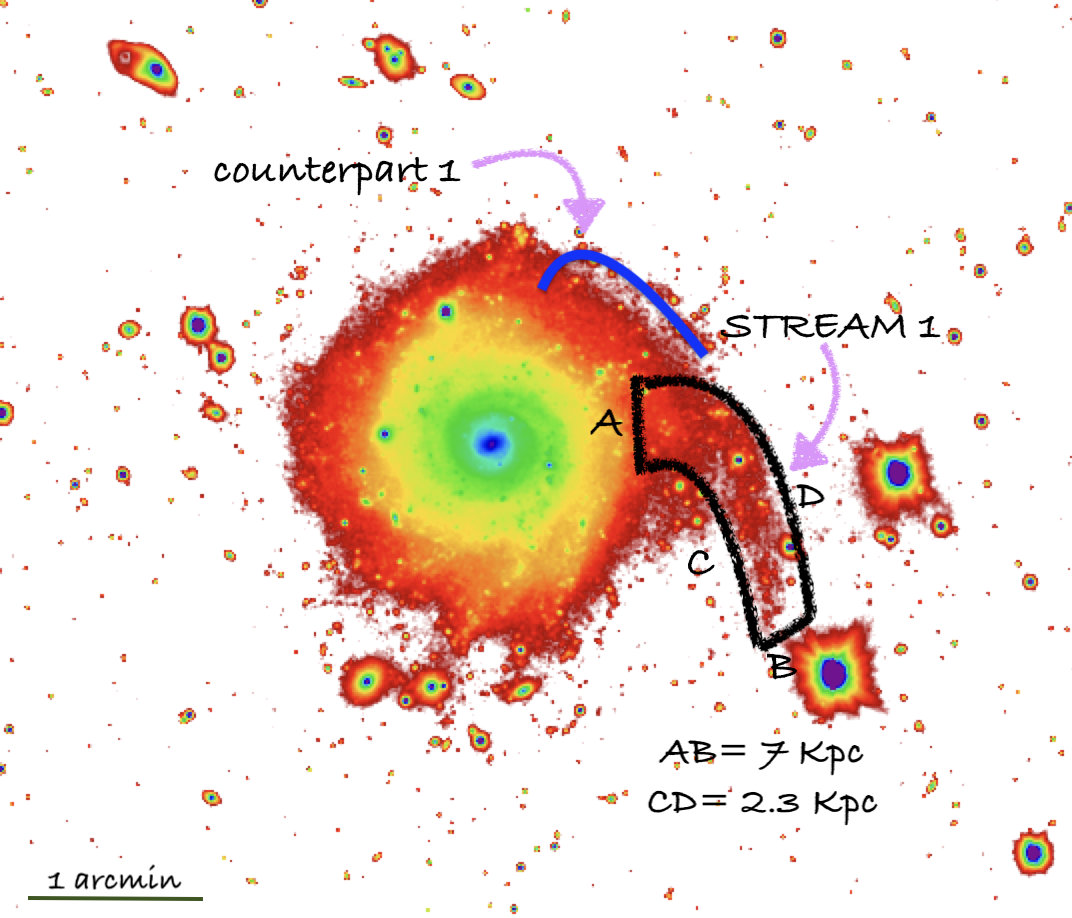}}
{\includegraphics[width=0.46\linewidth]{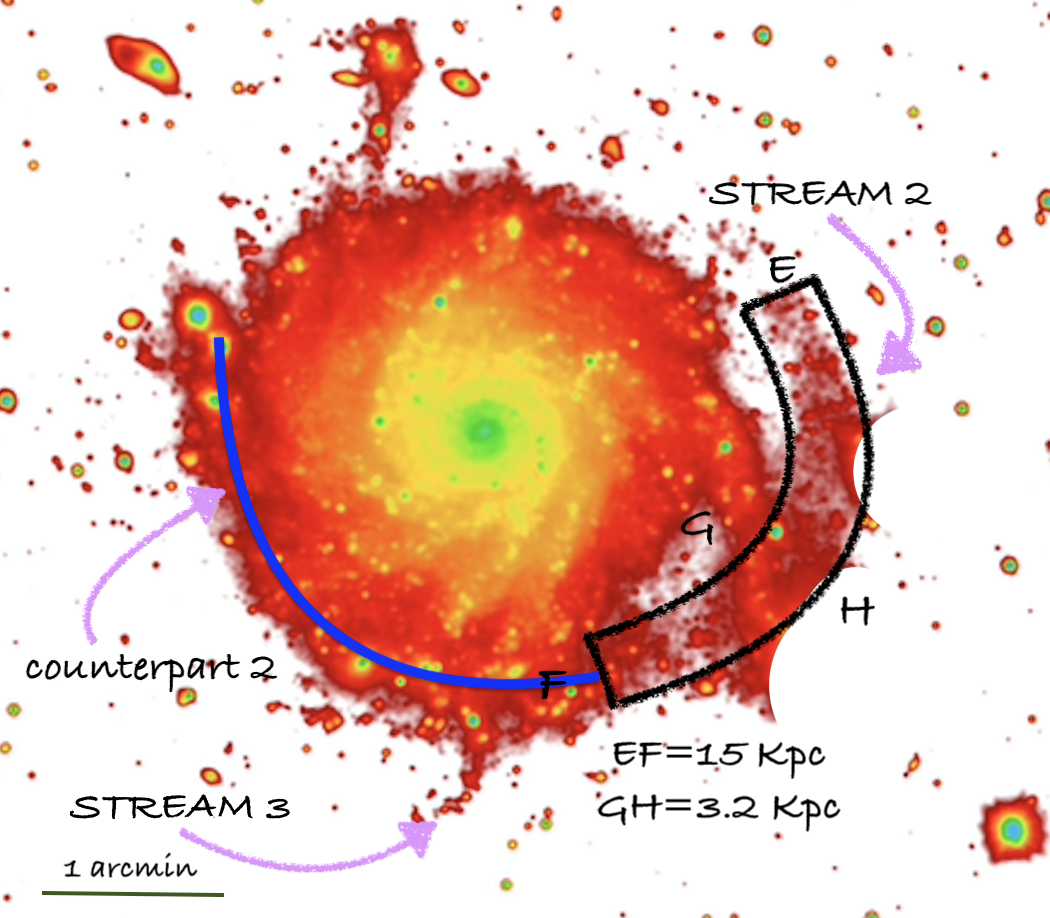}}
     \caption{Deep VST image in the $g$ band (\ros{in false color scale})  showing the outskirts of \ic5332\ with annotated stellar streams. North is up and East is to the left.
     Left panel: Stream 1, emerging to the west of the galaxy at a radial distance of $\sim$ 170 arcsec, is highlighted by the black polygon. 
     Right panel: Stream 2 and Stream 3, located to the southwest and west of the galaxy, respectively.
     Both images are smoothed by using a Gaussian kernel with $\sigma$ = 5 pixels. In both panels, with blue lines are indicated the regions corresponding to the counterparts.} 
    \label{fig:IC5332_streams}
\end{figure*}

\section{The western streams}\label{sec:stream}

The results in \Sec\ref{sec:SB_and_colour_profiles} provide a consistent multi-wavelength picture of \ic5332, revealing a complex structure shaped by both secular evolution and external accretion. The azimuthally averaged SB profiles, reaching $\mu_g \sim 29.1$, $\mu_r \sim 29.9$, and $\mu_i \sim 28.9$ mag arcsec$^{-2}$, show extended LSB outskirts deviating from a single exponential decline. A two-component Sérsic model is required to reproduce the profiles, with an extended outer component dominating in the $g$ and $r$ bands.

The wavelength-dependent structural parameters indicate older, metal-rich stellar populations in the central regions and a more diffuse, evolved component in the outskirts, consistent with an accreted stellar halo. The colour profiles display a negative inner gradient that inverts beyond $\sim$170 arcsec, where \ros{$g-r$ and $g-i$ become redder}—a signature of distinct stellar populations or accreted debris. Similar trends have been associated with external accretion or disc restructuring \citep[e.g.][]{bakos2008ApJ...683L.103B,laine2016A&A...596A..25L,Roediger2011MNRAS.413.2057R}. The converse behaviour of $r-i$, which becomes bluer at the same radii, remains unclear, although it may be driven by enhanced \ha\ emission.

At comparable radii, both PA and $\epsilon$ profiles change shape, suggesting that the outer isophotes are shaped by the geometry of accreted material rather than by the main stellar disc. The stellar mass surface density profiles reinforce this scenario, showing that a significant fraction of the mass lies in the outskirts, hosting old or metal-rich stars or with stronger \ha-\ntwo\ emission. All these indicators point to the presence of an extended, accreted stellar component in \ic5332, consistent with the detection of several LSB streams in its outer regions (see \Figs\ref{fig:composite_and_gband} and \ref{fig:IC5332_streams}), clearly visible in the deep VST imaging. Similar features have been observed in other nearby galaxies, confirming that such stellar streams are a common signature of past accretion events \citep[e.g.][]{sola2022A&A...662A.124S,pippert2025ApJ...980..244P}.

In particular, in the western direction, two major features can be identified. In the left panel, at a radial distance of $\sim $170 arcsec (i.e. $\sim 6.4$ kpc, $1.4 \, R_{\rm e, r}$), the first stellar stream begins to emerge prominently, and we highlight it as `Stream 1'. \ros{It seems to resemble a spiral arm}, since it follows a similar orientation to the other arms, but with a quite different pitch angle, which would suggest some interaction. Stream 1 appears brighter toward the galaxy, and, as shown in the left panel of Fig. \ref{fig:IC5332_streams}, it is clearly disconnected from the bright star located at its southern end. This supports the conclusion that the feature is not an imaging artifact caused by scattered light from the star. Its projected length is approximately 7 kpc, with an average width of about 2.3 kpc.  
A second feature, i.e., `Stream 2', is shown in the right panel. It is located at a slightly larger distance from the galactic center, at a radial distance of $\sim $ 220 arcsec (i.e. $\sim 7.6$ kpc, i.e. $1.7 \, R_{\rm e, r}$), that is the distance where the bump in the $r$-band SB profile emerges; it is instead a completely disconnected feature, which crosses the Stream 1, and does not follow the spiral arms pattern. It exhibits an arc-like shape oriented toward the center of \ic5332, which also makes it easily distinguishable from a possible stellar artifact in its vicinity. It extends over a projected length of 15 kpc and has an average width of about 3.2 kpc. 
In addition to these two main streams, the galaxy is characterized by smaller in size streams, for example `Stream 3', which are barely detectable due to their very LSB. 
These outer features are morphologically distinct, then suggesting a tidal origin, possibly the debris from a disrupted satellite galaxy \citep[e.g.,][]{laine2016A&A...596A..25L}. To confirm the distinct nature of Stream 1 and Stream 2, we computed the associated extinction-corrected average integrated colors, accounting also for the residual background fluctuations, and compared them with those of regions (named `counterpart 1' and `counterpart 2', respectively) located at the same radial distances from the center of the galaxy, as shown in \Fig \ref{fig:IC5332_streams}.
As previously mentioned, we focus solely on the $g - r$ color, which is the most informative for our purposes.
We find that Stream 2 and its Counterpart 2 exhibit $g - r$ values in the range 0.5–0.6 mag and 0.2–0.3 mag, respectively. Similarly, Stream 1 shows a $g - r$ color in the range 0.3–0.4 mag, whereas Counterpart 1 is bluer, with $g - r$ = 0.1–0.2 mag. The presence of such redder colours in the two streams is also supported by the rising radial profile measured within a wedge oriented toward the west. For both streams, their average colors are significantly redder than those of their respective counterparts, as expected in the case of accreted material surrounding a galaxy with the morphology of \ic5332.
In particular, the average color of Stream 2 is broadly consistent with the typical $g - r$ values reported for dwarf and LSB galaxies \citep{marleau21,LaMarca2022A&A...659A..92L, venhola22, Zaritsky2022ApJS..261...11Z}. This agreement suggests a possible scenario in which the stream's origin is associated with the complete or partial disruption of such an object, although other interpretations cannot be entirely excluded based on color data alone. The progenitor might still be present but obscured from view by one of the bright foreground stars located in that region.
In the case of Counterpart 1, it is not surprising that it appears bluer than Stream 1, given that it is located in a spiral arm of the galaxy.
All the findings listed above support the idea that \ic5332\ has undergone external accretion events that contributed to the buildup of its outer disc. Such processes are in line with the theoretical expectations for the inside-out growth of spiral galaxies, where the central regions form early and evolve through secular processes, while the outer structures are gradually assembled via accretion of satellites and diffuse material \citep[e.g.,][]{pillepich2015ApJ...799..184P,Rodriguez-Gomez+16_Illustris, Grand+17_Auriga}.
To further exclude that the detected streams are artifacts caused by scattered light from nearby bright stars, we compared our VST images with independent optical data from the DESI Legacy Surveys DR10. Both streams are clearly detected also in the DESI data, confirming their physical nature. This comparison is discussed in App. \ref{app:desi_comparison}.

\section{Final remarks}\label{sec:discussion_conclusione}

The VST-SMASH survey \citep{Tortora+24_VST-SMASH} has been designed to detect low surface brightness features in the outskirts of a distance-limited sample of nearby late-type galaxies, within a distance of $D < 11$ Mpc. This is a largely unexplored regime in extragalactic astronomy, mainly due to the observational challenges posed by the detection of diffuse structures at very faint levels. Such a goal requires both multi-band, deep and wide-field imaging capabilities, which are provided by the VST in the optical $g$, $r$, and $i$ bands. Notably, the VST-SMASH survey also offers the unique opportunity to serve as the optical counterpart to the {\it Euclid} Wide Survey footprint, providing complementary ground-based data for future multi-wavelength studies (\citealt{EuclidWide2022,hunt2025A&A...697A...9H}).

In this first presentation paper, we focused on the galaxy \ic5332\ to illustrate the depth, quality, and scientific potential of the VST-SMASH dataset. We reach $1\sigma$ limiting surface brightnesses, measured within an aperture of $100\, \rm arcsec^{2}$, of $\sim 30.7$, $30.5$, and $29.5$ mag arcsec$^{-2}$, with PSF FWHM values of $\sim 0.9$, $1.1$, and $0.8$ arcsec in the $g$, $r$, and $i$ bands, respectively. These results confirm that the survey achieves the required depths, reaching in the $g$ and $r$ bands levels comparable to those of the {\it Euclid} Wide Survey \citep{hunt2025A&A...697A...9H}.
Through detailed surface photometry, we derived surface brightness and color profiles down to unprecedented depths, revealing structural components and gradients not detectable in shallower surveys. We performed double-component Sérsic decompositions in all bands, examined the radial variation of structural parameters, and investigated stellar population gradients through both color and mass profiles. Crucially, we also identified and characterized faint stellar streams in the galaxy’s outskirts, interpreting them as evidence of ongoing or past accretion events.
While a single-component model fails to capture the full photometric structure, \ic5332\ is well fitted by a double-component Sérsic profile. The inner component is consistent with previous literature results, while the more extended component appears larger in the $r$ band. The galaxy is characterized by a negative $g-i$ colour gradient within one effective radius, which, even after accounting for internal extinction, is somewhat steeper than the average for galaxies of similar mass \citep{Tortora+10CG}, and in agreement with the steep gas-phase metallicity gradients reported by \citet{Groves+23_PHANGS-MUSE}. These findings are consistent with a dissipative monolithic collapse scenario, reinforced by supernova-driven winds. In the outer regions, the scenario is more complex: distorted spiral arms and stellar streams (e.g., Streams 1 and 2 in \Fig\ref{fig:IC5332_streams}) complicate the structure. These streams present integrated colors consistent with disrupted dwarf and/or low surface brightness satellites. When using azimuthally averaged colour profiles, we also observe opposite trends in the $g - r$ and $r - i$ colour.  Their nature remains puzzling; we suggest that this behaviour may be driven by a stronger contribution from emission lines in the $r$-band filter, an interpretation that will require further investigation.

Although this study focuses on a single galaxy, it clearly demonstrates the exceptional potential of the VST-SMASH dataset to uncover and characterize faint features in the outskirts of spiral galaxies. With this work, we lay the groundwork for future analyses of the remaining targets in our survey, opening a new window into the low surface brightness universe in the nearby cosmos.
In future studies, we will extend this analysis to the full VST-SMASH sample, aiming to perform a systematic morphological classification of stellar streams and diffuse substructures, search for dwarf galaxies and globular clusters, investigate their photometric properties, and quantify their occurrence rates. By combining this information with ancillary multi-wavelength data, including space-based {\it Euclid} imaging, and comparisons to predictions from cosmological simulations, we aim to place robust constraints on the role of hierarchical accretion in shaping the outskirts of nearby spiral galaxies.

\bibliography{smash_biblo}

\begin{appendix} 

\section{Surface photometry procedure}\label{app:SB_photometry_details}

In this section, we outline the main steps of the data analysis used to derive surface brightness profiles:

 i)  Cropping: the first step of the data analysis consists of cropping the image edges \ros{ (i.e., regions with a low S/N)} to retain a sufficiently large area that encompasses the faintest light surrounding the target.

 ii) Masking: on the cropped image, we perform then meticulous manual masking of all potential contaminants in the images, including foreground and background galaxies, stars, and galactic cirrus. This step ensured that only the target structures were analyzed while excluding sources of interference.

 iii) Correcting for scattered light from bright stars: to mitigate the effects of scattered light from bright stars present in the field, a modeling and subtraction procedure was applied. Bright stars in the field were modeled using a 2D profile and subsequently subtracted from the original image out to large radii. However, this analytical modeling often leaves systematic residuals, primarily due to the presence of non-symmetric features such as optical 'ghosts'—internal reflections typical of bright sources in these datasets.
 They break the expected symmetry of the source and cannot be perfectly removed by the model. Consequently, these residual regions were manually masked to ensure the integrity of the LSB measurements. These masks are visible as white circular patches in the right panel of Fig. \ref{fig:IC5332_streams}. This correction minimizes the influence of stellar light on the LSB measurements, preserving the integrity of the data.

 iv) Assessing the limiting radius and background fluctuations: following the removal of contaminating stars, the limiting radius of photometry ($R_{lim}$) was determined. This radius corresponds to the point at which the galaxy's light merges with residual background fluctuations. To estimate $R_{\rm lim}$, the light distribution was analyzed in circular annuli centered on the galaxy, to sample the residual sky without being biased by the morphology of the galaxy’s isophotes.  As the images were pre-processed for sky subtraction, the residual background level was found to be near zero. The average residual value and its RMS are then used in step v), contributing to the estimate of the surface brightness and its associated errors.

 v) Isophotal fitting: we performed the isophotal analysis of the galaxy light distribution using the Ellipse module from the photutils python package. The fitting procedure was carried out independently on each optical band ($g$, $r$, $i$), adopting a fixed center across all filters to ensure consistency in the structural comparison.  In each band, we determined the galaxy center by first applying a Gaussian smoothing with a 3-pixel kernel to the image, and then identifying the brightest central region of 9 square pixels. The central pixel of this region was adopted as the galaxy center. The initial input parameters — position angle (PA) and ellipticity ($\epsilon$, defined as $\epsilon = 1 - (b/a)$, where a and b are the semi-major and semi-minor axes, respectively)— were set based on literature values, providing a physically motivated starting point for the isophotal fitting. The algorithm proceeds by iteratively adjusting the semi-major axis (sma) with a step of 0.1 pixels, while solving for the best-fitting isophotes at increasing radii, up to the edges of the frames. The center was kept fixed (fix$\_$center=True), and the fit was performed using median sampling along the isophotes (integrmode='median'). The final isophotal solution was derived in a non-linear mode (linear=False), with convergence controlled by a maximum number of iterations and a tolerance on gradient errors, which for the standard setup are maxit=100 and maxgerr=50, respectively. This approach allowed us to extract reliable radial profiles of surface brightness, ellipticity, and position angle, which are then used to investigate the structural components and stellar population gradients in the galaxy.

These steps ensured the robust analysis of LSB components, enabling a deeper understanding of their properties and their connection to the broader astrophysical environment.

\section{Limiting surface brightness calculation}\label{app:limiting_SB_calculations}

In this study, we calculate the limiting surface brightness $\mu_{\rm lim}$
within a 100 arcsec$^2$ region as a function of the signal standard
deviation $\sigma$ based on the methods outlined in
\citet{hunt2025A&A...697A...9H}. We use two complementary approaches:

-- NoiseChisel with gnuastro. We use the
NoiseChisel software \citep{akhlaghi15}, which is specifically
designed to detect highly extended, faint objects embedded in noisy
backgrounds. NoiseChisel runs on tiles of 100 arcsec$^2$ and calculates
the median $\sigma$ in regions where no detections are present,
isolating the empty sky background. These results are shown in
Table~\ref{tab:sky} as line (1) in each band.

-- Gaussian fitting on masked data. Following
\citet{roman20}, we fit a Gaussian to the distribution of the masked
sky-background signal using masks from the NoiseChisel detections (as in
method 1). The Gaussian best-fit standard deviation ($\sigma$) across
the masked background image is then used to calculate limiting AB
magnitudes for each band, as shown in line (2) of Table~\ref{tab:sky}.

These two methods probe background variations on different spatial
scales: method (1) measures the local pixel-to-pixel noise, while method
(2) is sensitive to large-scale background fluctuations across the
masked image. As shown in Tab.\ref{tab:sky}, both methods yield consistent
values, differing by less than 0.1 mag.

We stress that to assess
the effective depth at which reliable SB measurements can be performed,
we additionally quantify the large-scale residual background
fluctuations by analysing the radial count profiles and identifying the
plateau region where the profiles flatten and become dominated by
background residuals, as described in step iv) of App. \ref{app:SB_photometry_details}.
 These large-scale background fluctuations are
used both to define the effective SB depth adopted in the analysis of
the radial profiles and to build the corresponding uncertainty budget.

\begin{table}[t!]
\setlength{\tabcolsep}{3.6pt}
\centering
\caption[]{SB limiting magnitudes \mulim\ within 100\,arcsec$^{2}$ areas in empty sky regions.}\label{tab:sky}
\centering
\begin{threeparttable}
\begin{tabular}{lccc}
\hline
\hline
\noalign{\vskip 3pt}
\multicolumn{1}{c}{Method}  & Depth  \\

&[mag arcsec$^{-2}$]\\
\hline
\\ 
\multicolumn{1}{c}{(1) $\mu_{g}$ 1$\sigma$ SB limit}  & 30.7 \\
\multicolumn{1}{c}{(2) $\mu_{g}$ 1$\sigma$ SB limit} & 30.7 \\
\hline
\multicolumn{1}{c}{(1) $\mu_{r}$ 1$\sigma$ SB limit}  & 30.5 \\
\multicolumn{1}{c}{(2) $\mu_{r}$ 1$\sigma$ SB limit} & 30.4 \\
\hline
\multicolumn{1}{c}{(1) $\mu_{i}$ 1$\sigma$ SB limit}  & 29.5 \\
\multicolumn{1}{c}{(2) $\mu_{i}$ 1$\sigma$ SB limit}  & 29.5 \\
\noalign{\vskip 1pt}
\hline
\end{tabular}
\begin{tablenotes}
        \footnotesize
        \item Each line corresponds to the different approaches: (1) the standard deviation among 100\,arcsec$^2$ tiles following gnuastro/noisechisel; and (2) Gaussian fits of the masked sky regions following \citet{roman20}.
    \end{tablenotes}
\end{threeparttable}
\end{table}

The achieved formal depths—30.7, 30.5, and 29.5 mag arcsec$^{-2}$ in the
$g$, $r$, and $i$ bands—match the planned depths of VST–SMASH
\citep{Tortora+24_VST-SMASH}. The effective SB depth reached by the
azimuthally averaged profiles is discussed in Sect.~4.1. This approach
follows closely the methodology adopted in
\citet{ragusa2021A&A...651A..39R} and ensures a conservative and
physically meaningful assessment of the SB limits.

\section{One-component Sérsic fit}\label{app:sersic1comp}

\begin{figure}
    \centering
    \includegraphics[width=0.84\linewidth]{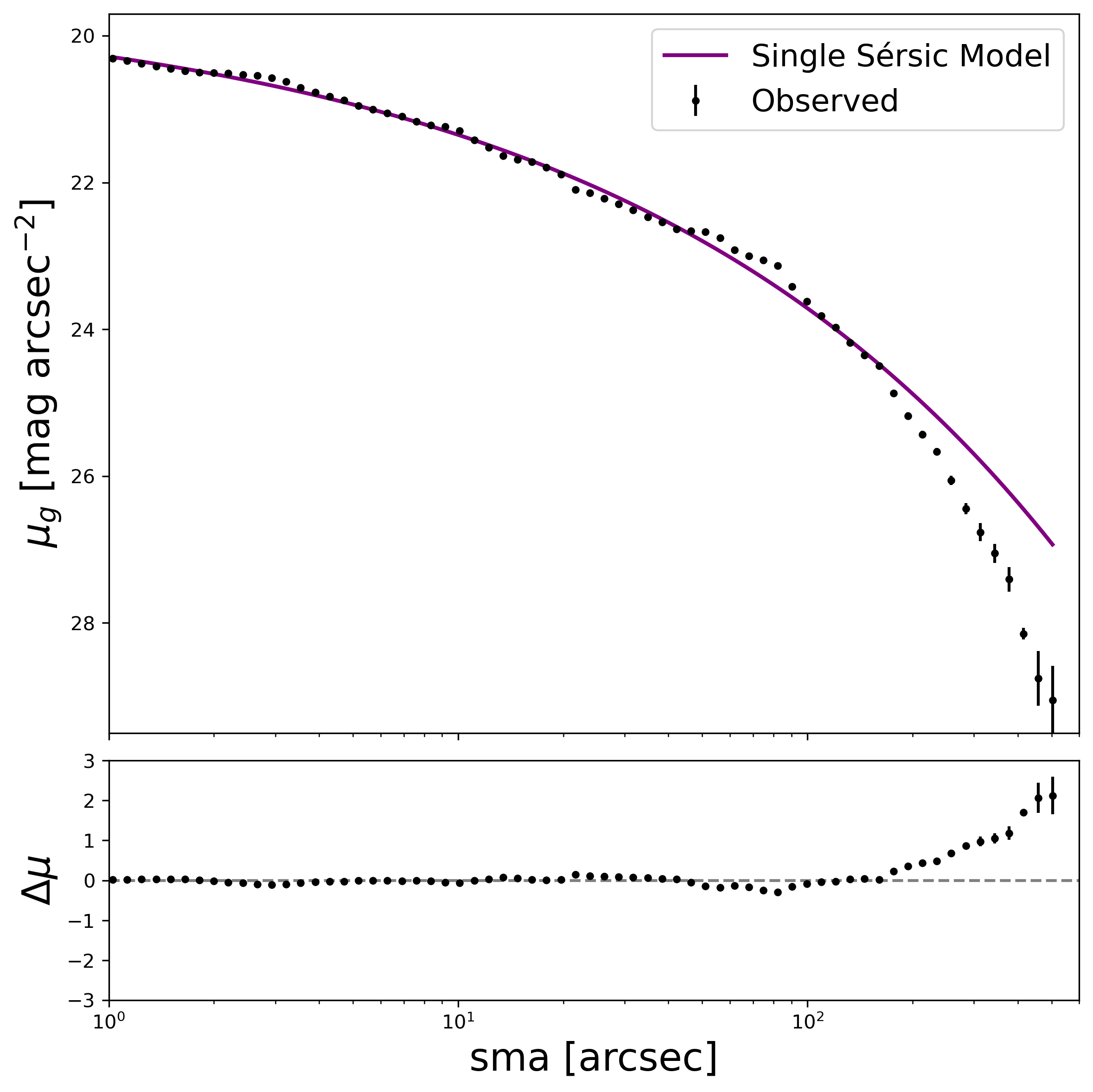}
    \includegraphics[width=0.84\linewidth]{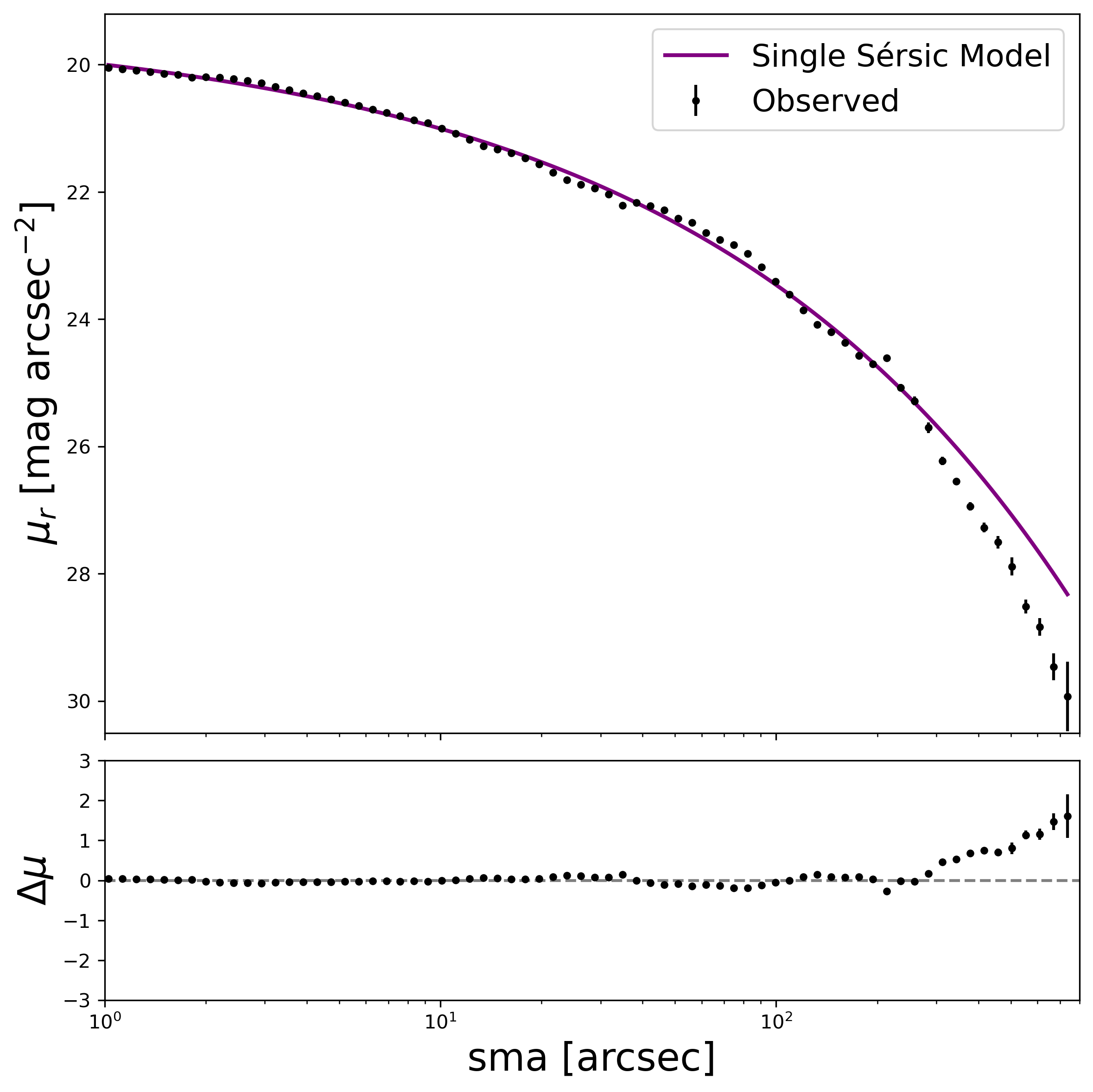}
    \includegraphics[width=0.84\linewidth]{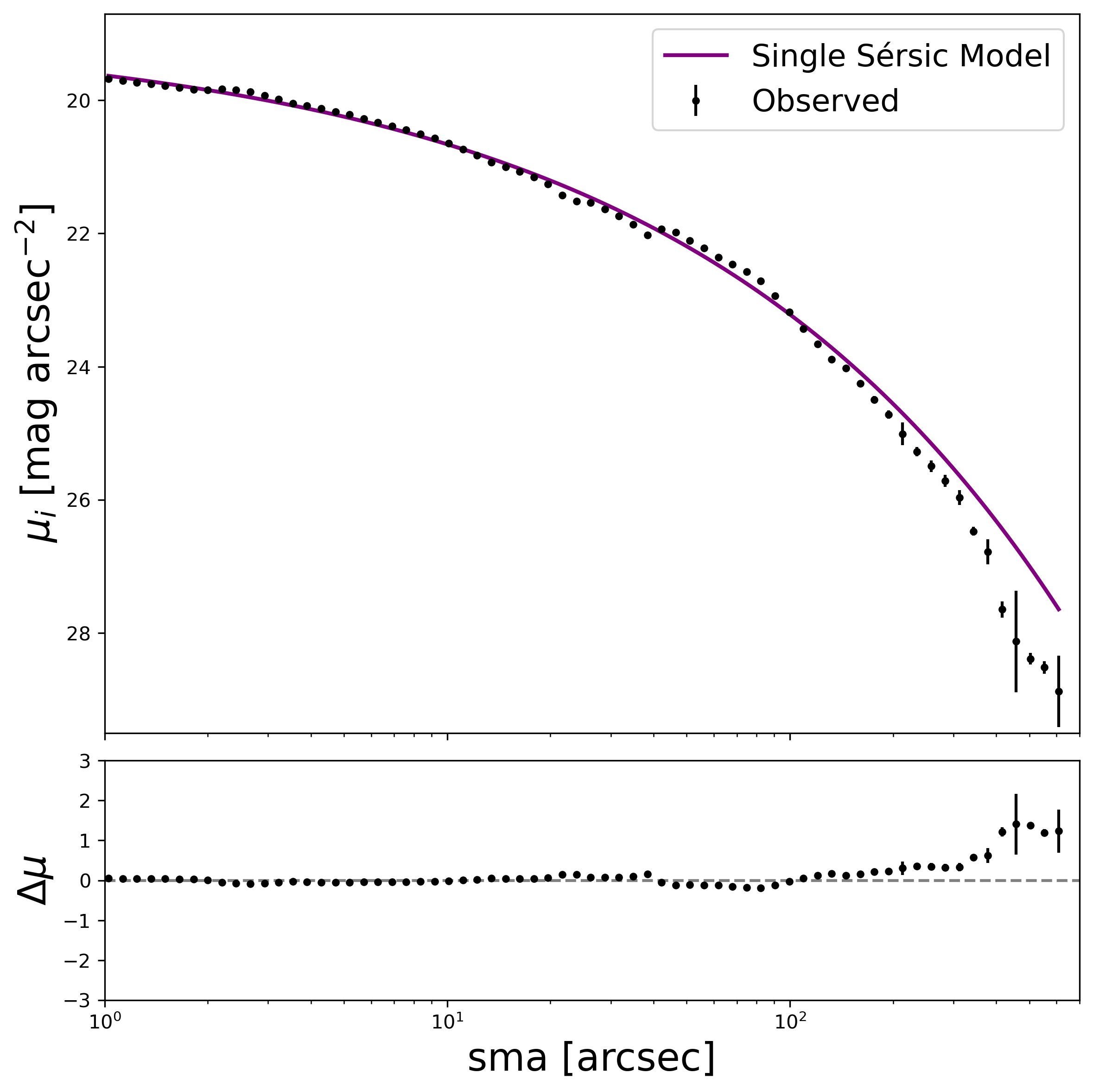}
    \caption{Sérsic fit of \ic5332. From the top to the bottom: One-component model of the azimuthally-averaged surface brightness profiles of \ic5332\ in the $g$, $r$, and $i$ bands, respectively. In each plot, the violet line indicates the best-fit model of the observed light profile in black. The bottom panels show the residuals of the fit, $\Delta\mu = \mu_{obs} - \mu_{model}$.}
    \label{fig:1comp_sersic}
\end{figure}   

In this appendix, we present the results of the one-component Sérsic fits performed on the azimuthally averaged surface brightness profiles of \ic5332\ in the $g$, $r$, and $i$ bands. \Fig \ref{fig:1comp_sersic} shows, from top to bottom, the best-fit Sérsic models (red lines) overlaid on the observed profiles in each band. 

As discussed in \Sec\ref{subsec:decomp}, although the single-component model provides a reasonable approximation in the inner regions, it systematically fails to reproduce the outer light distribution in all three bands.
Indeed, in all bands, the one-component Sérsic model yields a reduced $\chi^2$ values (62.420, 69.27 and 90.81 in the $g$, $r$ and $i$ bands) and a sum of the absolute residuals (17.31, 10.41, 10.07 mag in the $g$, $r$ and $i$ bands) higher than the 
two-components fit (see \Sec \ref{app:sersic2comp_app}).
The improvement provided by the two-component model is also 
clearly visible in the residual profile in the bottom panels of Fig. \ref{app:sersic2comp_app} when we compare them with the one in the one-component model, shown in the bottom panels of Fig. \ref{app:sersic1comp}.

\section{Two-component Sérsic fit}\label{app:sersic2comp_app}

\begin{figure}
    \centering
    \includegraphics[width=1\linewidth]{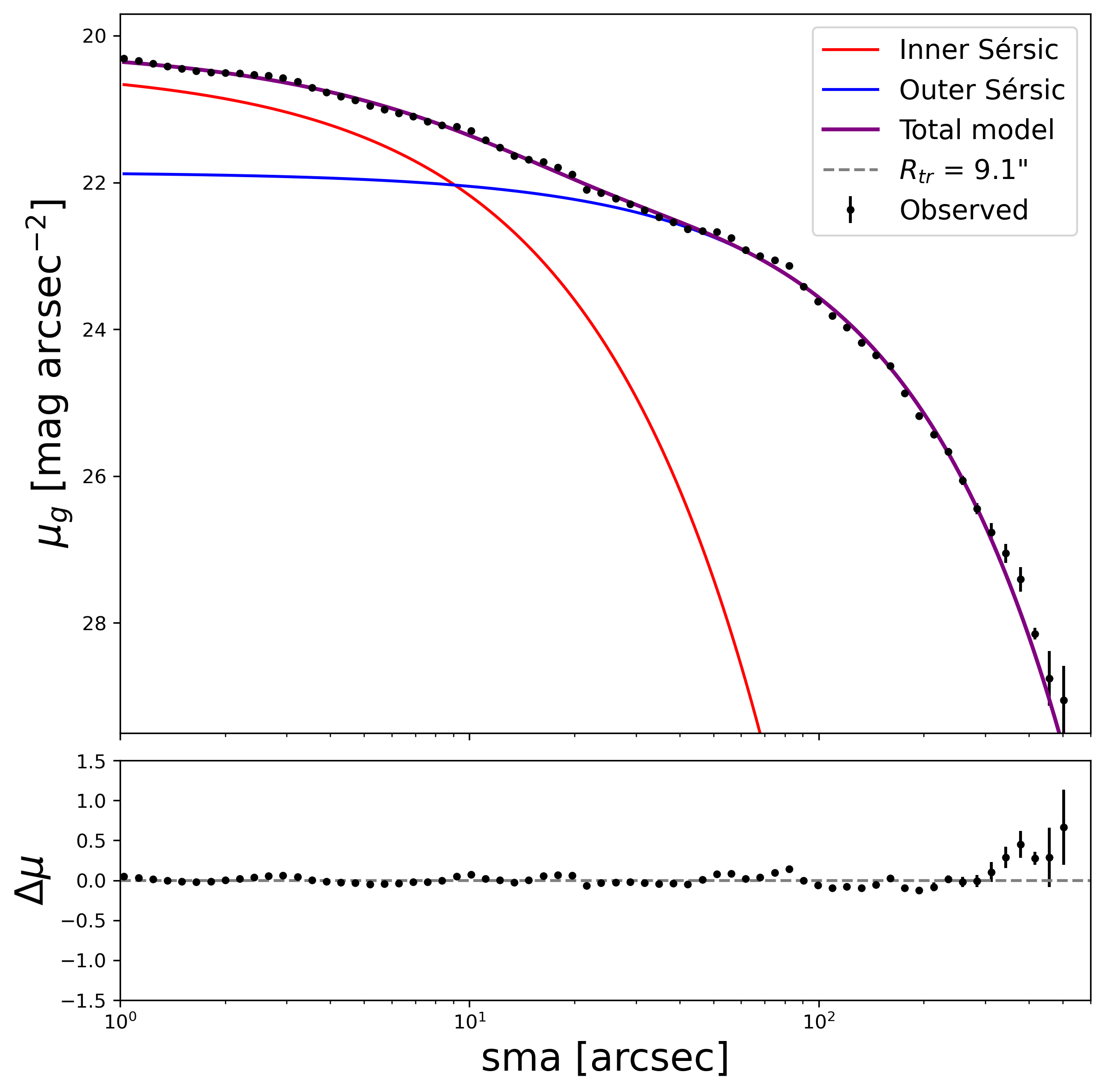}
    \includegraphics[width=1\linewidth]{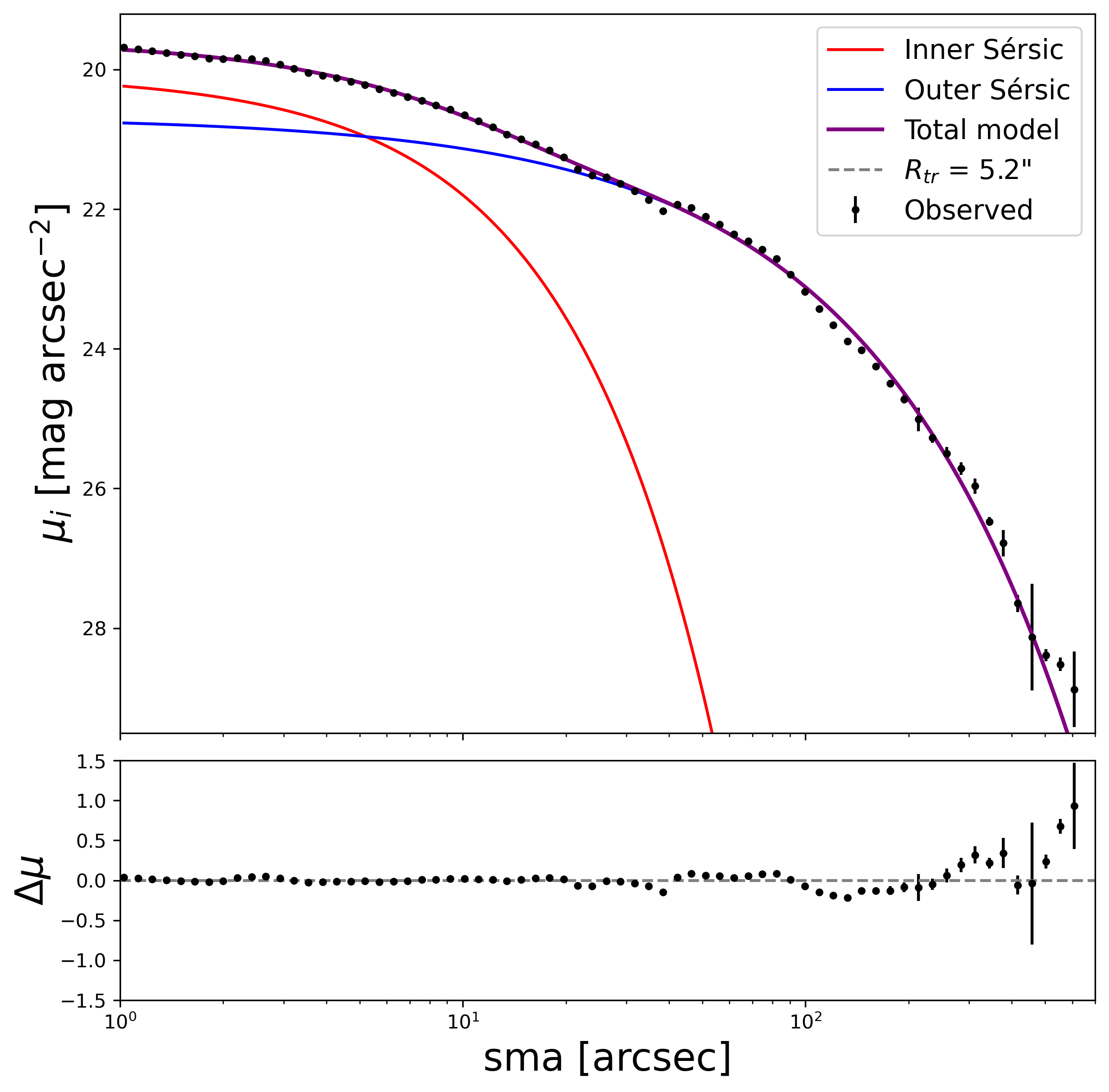}
    \caption{Two-component Sérsic fit of the azimuthally-averaged surface brightness profiles of \ic5332, in the $g$ (on the top) and $i$ (on the bottom) bands. In both panels, the red line indicates the best-fit model of the observed light profile. The bottom panels show the residuals of the fit, $\Delta\mu = \mu_{obs} - \mu_{model}$.}
    \label{fig:2comp_sersic}
\end{figure}

In this Section, we show in Fig. \ref{fig:2comp_sersic} the results of the two-component Sérsic fits performed on the azimuthally averaged surface brightness profiles of IC 5332 in the $g$ and  $i$ bands.

The goodness of fit is quantified using the reduced $\chi^2$, 
allowing a direct comparison between single- and two-component 
Sérsic models. However, we note that in deep surface brightness 
profile analyses the reduced $\chi^2$ values can be systematically 
larger than unity due to correlated uncertainties between adjacent 
radial bins and residual background systematics that are not fully 
captured by the formal photometric errors. For this reason, the 
reduced $\chi^2$ should be interpreted as a relative indicator of goodness of fit rather than as an absolute statistical estimator.
In addition to the reduced $\chi^2$, we therefore quantify the 
goodness of fit using the sum of the absolute residuals, 
$\mathrm{sum}(|\Delta\mu|)$, where 
$\Delta\mu = \mu_{\rm obs} - \mu_{\rm model}$. 
This metric provides a robust measure of the typical deviation between the data and the model.
In all bands, the two-component Sérsic model yields both lower 
reduced $\chi^2$ values (14.67, 23.70, 27.11 in the $g$, $r$ and $i$ bands) and sum of the absolute residuals (1.068, 1.012, 1.954 mag in the $g$, $r$ and $i$ bands) than the 
single-component fit (see \Sec \ref{app:sersic1comp}), confirming that it provides a statistically 
better description of the observed profiles.
This motivates the adoption of a more flexible two-component model, as described in the main text.

\section{Impact of Bright Stars on the Radial SB Profiles}
\label{app:quadrants}
\begin{figure}
    \centering
    \includegraphics[width=\linewidth]{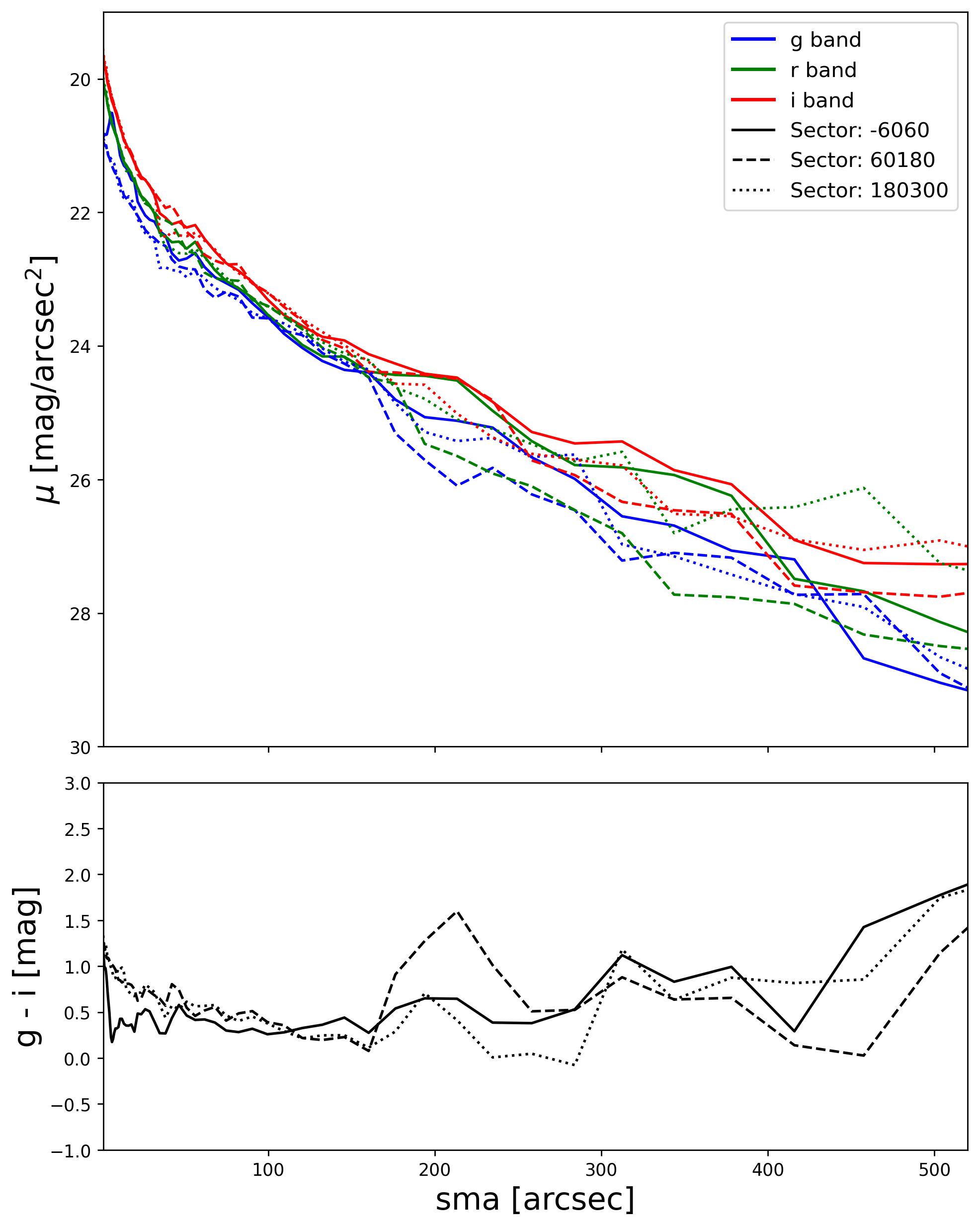}
    \caption{Azimuthally-averaged surface
brightness profiles of IC 5332 in VST g (blue), r (green) and i (red)
bands, in three different angular sector, anticlockwise with respect to the horizontal: (-60\textdegree ; 60\textdegree),  (60\textdegree ; 180\textdegree), (180\textdegree ; 300\textdegree), represented with the solid, dashed and dotted lines, respectively. 
The g − i (in black) color profiles of the angular sectors are 
also shown in the bottom panel.}
    \label{fig:sectors}
\end{figure}

The presence of two bright stars at the south-west edge of IC~5332 raises the possibility that diffuse PSF halos could contaminate the surface brightness measurements in that region.

To quantitatively assess this effect and exclude any bias in the radial SB profiles, we performed a sector-based analysis.
We divided the galaxy into three angular sectors (-60\textdegree ; 60\textdegree),  (60\textdegree ; 180\textdegree) and (180\textdegree ; 300\textdegree), with respect to the horizontal. We independently derived the radial SB profiles in each sector using the same isophotal fitting procedure described in Sect. \ref{app:SB_photometry_details}. One of these sectors encompasses the entire region potentially affected by scattered light from the bright stars (-60\textdegree , 60\textdegree) and overlaps with the area where the stellar streams are detected.
Figure~\ref{fig:sectors} shows the SB profiles obtained for the three sectors in the $g$, $r$, and $i$ bands. The sector containing the bright stars does not show any systematic excess or deviation with respect to the others.
This test demonstrates that the contribution of stellar PSF halos to the radial SB profiles is negligible and does not affect the azimuthally averaged profiles adopted in the main analysis. We therefore conclude that the scientific results based on the SB profiles are robust against potential contamination from nearby bright stars.

\section{Comparison with DESI Legacy Surveys DR10 Imaging}
\label{app:desi_comparison}

\begin{figure}
    \centering
    \includegraphics[width=\linewidth]{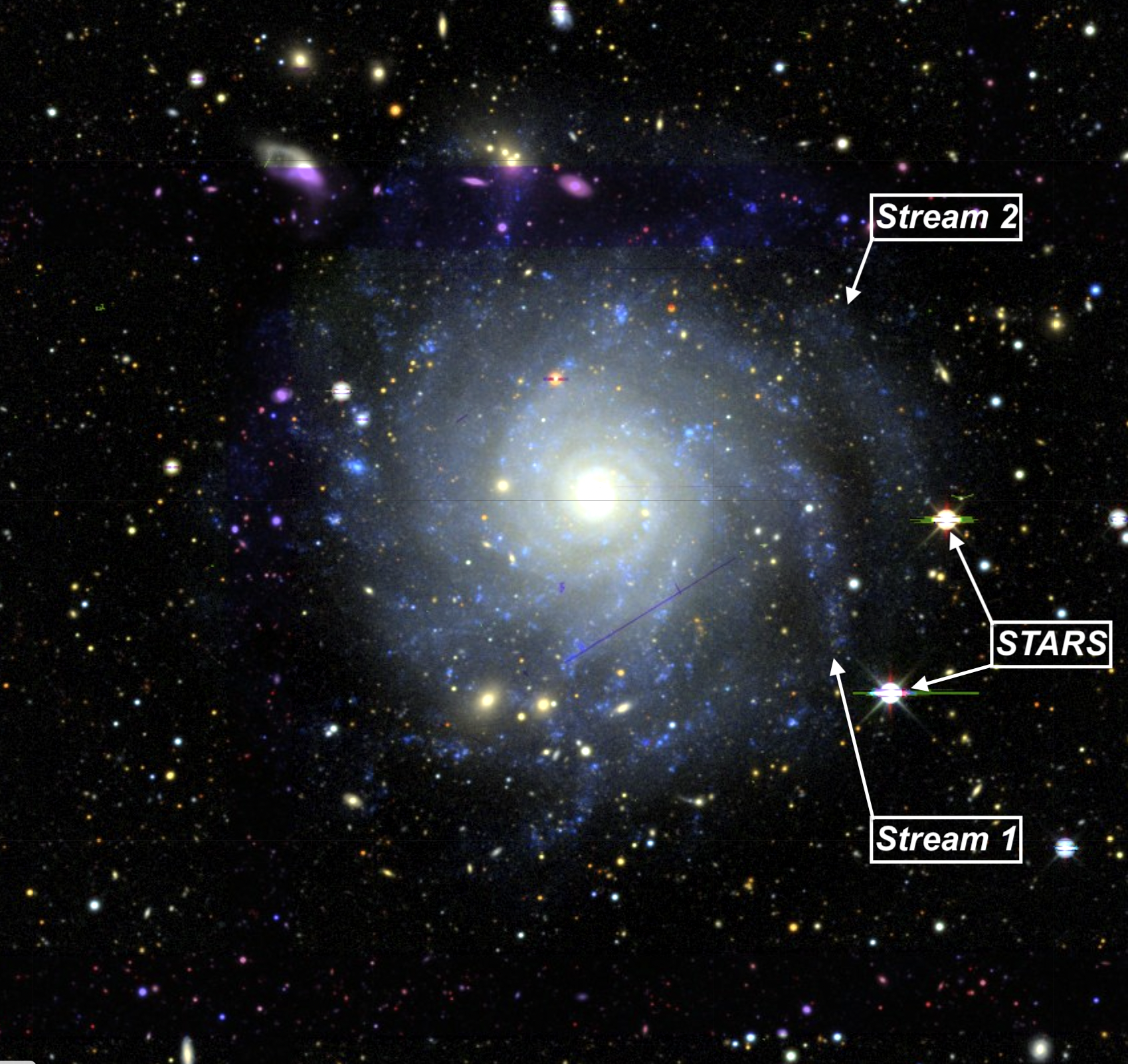}
    \caption{The DESI DR10 image of IC 5332. The stars and streams are labeled in white.}
    \label{fig:desi}
\end{figure}

To further verify the reality of the low-surface-brightness streams detected in the outskirts of IC~5332, and to exclude the possibility that they are artifacts caused by scattered light from nearby bright stars, we compared our VST images with independent optical imaging from the DESI Legacy Surveys DR10.
The DESI DR10 image of IC 5332 in \Fig \ref{fig:desi} show significantly reduced scattered light from adjacent bright stars in the region of interest. 
Unfortunately, the data from DESI DR10 suffer from many artifacts, especially regarding the coverage of the external parts, which prevents us from being able to use them consistently for our purposes.
However, both stellar streams discussed in \Sec\ref{sec:stream}, including stream~2, which partially overlaps with the region affected by the bright stars in the VST data, are clearly visible also in the DESI images. This confirms that these features are genuine stellar structures associated with IC~5332 and not artifacts produced by stellar PSF halos or background subtraction residuals.

The detection of the streams in two independent datasets significantly strengthens the robustness of our analysis and supports their interpretation as accretion-related structures.

\section{Impact of the extended PSF wings}
\label{app:psf}

\begin{figure}
    \centering
    \includegraphics[width=\linewidth]{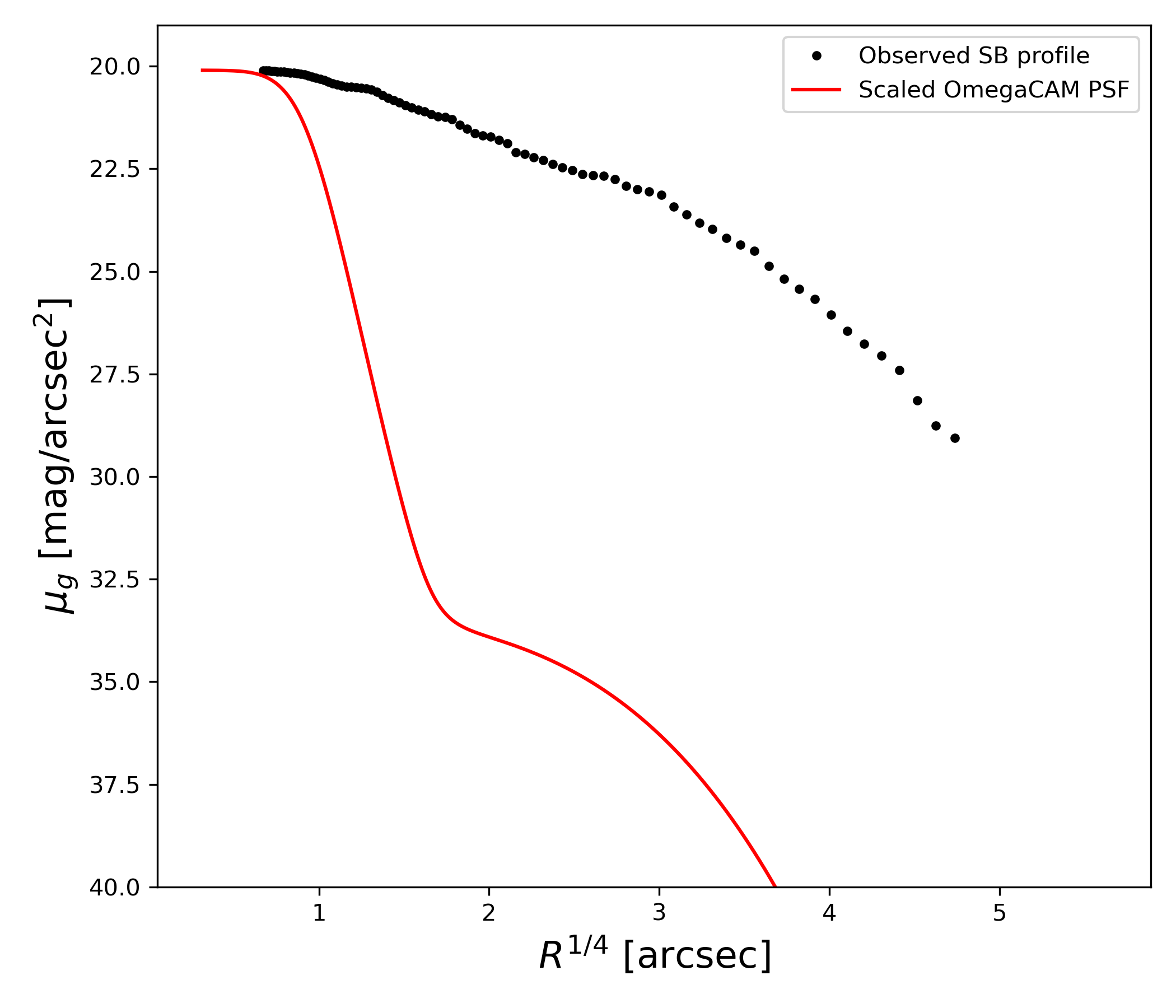}
    \caption{Comparison between the observed $g$-band surface brightness profile of IC 5332 (black dots) and the scaled OmegaCAM PSF model (red line). The PSF model accounts for both the core (Moffat) and the extended exponential wings as described in Capaccioli et al. (2015). }
    \label{fig:psf}
\end{figure}

OmegaCAM is known to exhibit extended, exponential-like PSF wings, as detailed by \citet{Capaccioli2015A&A...581A..10C}. To assess their impact on our analysis, we reconstructed a composite PSF model—comprising a Moffat core and an exponential outer component—and scaled it to match the central surface brightness of the galaxy. As shown in Fig. \ref{fig:psf}, the scaled PSF profile, in red, remains several magnitudes fainter than the observed galaxy emission, in black, at radii larger than $\sim 150$ arcsec. Even adopting the conservative model with extended exponential wings as in Fig. B1 from \citet{Capaccioli2015A&A...581A..10C}, the PSF contribution stays well below the measured surface brightness in the outer regions. This demonstrates that the extended wings do not significantly affect our surface brightness profiles or the detection of the stream-like structure.

\end{appendix}
\begin{acknowledgements}
R.R. acknowledges financial support grants through INAF-WEAVE StePS funds and through PRIN-MIUR 2020SKSTHZ.
MS acknowledges the support by the Italian Ministry for Education, University and Research (MIUR) grant PRIN 2022 2022383WFT “SUNRISE”, CUP C53D23000850006, and by VST funds.
Based on data collected with the INAF VST telescope at the ESO Paranal Observatory.
\end{acknowledgements}

\end{document}